\documentclass{cernyrep}
\usepackage{texnames}
\usepackage[T1]{fontenc}
\usepackage{varwidth}
\usepackage{xcolor}

\begin{document}
\title{Mathematical and Numerical Methods for Non-linear Beam Dynamics}

\author{W. Herr}

\institute{CERN, Geneva, Switzerland}

\maketitle 

\begin{abstract}
Non-linear effects in accelerator physics are  important for both successful
operation of accelerators and during the design stage.
Since both of these aspects are closely related, they will be treated
together in this overview.
Some of the most important aspects are well described by methods
established in other areas of physics and mathematics.
The treatment will be focused on
the problems in accelerators used for particle physics experiments.
Although the main emphasis will be on accelerator physics issues,
some of the aspects of more general interest will be discussed.
In particular, we demonstrate that in recent years a framework has
been built to handle the complex problems in a consistent form,
technically superior and conceptually simpler than the traditional
techniques.
The need to understand the stability of particle beams has substantially
contributed to the development of new techniques and is an important
source of examples which can be verified experimentally.
Unfortunately, the documentation of these developments is often
poor or even unpublished, in many cases only available as lectures or
conference proceedings.
\end{abstract}

\section{Introduction and motivation}
\subsection{Single-particle dynamics}
The concepts developed here are used to describe single-particle
transverse dynamics in rings, i.e. circular accelerators or storage rings.
In the case of linear betatron motion the theory is rather complete
and the standard treatment \cite{bib:cs01} suffices to describe the
dynamics. In parallel with this theory the well-known concepts such
as closed orbit and Twiss parameters are introduced and emerge
automatically from the Courant--Snyder formalism \cite{bib:cs01}.
The formalism and applications are found in many textbooks (e.g. \cite{bib:wiedemann,bib:chaotig}).

In many new accelerators or storage rings (e.g.\ the Large Hadron Collider (LHC))
the description
of the machine with a linear formalism becomes insufficient and
the linear theory must be extended to treat non-linear effects.
The stability and confinement of the particles is not given a priori and
should rather emerge from the analysis.
Non-linear effects are a main source of performance limitations in such machines.
Non-linear dynamics has to be used in many applications and the basics are described
in the literature~\cite{bib:poin01, bib:lich01}.
A reliable treatment is required and the progress in recent years allows us
to evaluate these effects.
A very useful overview and details can be found
in \cite{bib:forest, bib:chao, bib:dragt}.

In this article, we restrict the treatment to the beam dynamics in rings, since
properties such as stability of particles restrict the form of the
tools to describe them.
However, many of the concepts (e.g.\ Lie transformations) can be very beneficial
for the analysis of single-pass systems.

\section{New concepts}
The key to the more modern approach shown in this article is to avoid the
prejudices about the stability and other properties of the ring.
Instead, we must describe the machine in terms of the objects it consists of with all
their properties, including the non-linear elements.
The analysis will reveal the properties of the particles such as, e.g., stability.
In the simplest case, the ring is made of individual machine elements such as magnets
which have an existence on their own, i.e.\ the interaction of a particle with a given element
is independent of the motion in the rest of the machine.
To successfully study single-particle dynamics, we must be able to describe the action
of the machine element on the particle as well as the effect of the machine element
in the presence of all other magnets in the ring~\cite{bib:herr}.

\subsection{Map-based techniques}
In the standard approach to single-particle dynamics in rings, the equations of motion
are introduced together with an ansatz to solve these equations.
In the case of linear motion, this ansatz is known as the Courant--Snyder treatment.
However, this treatment must assume that the motion of
a particle in the ring is stable and confined.
For a non-linear system this is a priori not known and the attempt to find a complete
description of the particle motion must fail.

In a more modern approach, we do not attempt to solve such an overall equation but rather
consider the fundamental objects of an accelerator, i.e.\ the machine elements themselves.
These elements, e.g.\ magnets or other beam elements, are the basic building blocks of
the machine.
All elements have a well-defined action on a particle which can be described independently
of other elements or concepts such as closed orbit or $\beta$-functions.
Mathematically, they provide a `map' from one face of a building block to the other,
i.e.\ a description of how the particles move inside and between elements.
In this context, a map can be anything from linear matrices to high-order integration
routines.

The collection of all machine elements makes up the ring and it is the combination
of the associated maps which is necessary for the description and analysis of the
physical phenomena in the accelerator ring.

The most interesting map is the one which describes the motion once around the machine,
the so-called one-turn map (OTM).
It contains all necessary information on stability and existence of closed orbit and
optical parameters.
The reader is assumed to be familiar with this concept in the case of linear beam
dynamics~\cite{bib:holzer}, where all maps are matrices and the Courant--Snyder analysis
of the corresponding OTM produces the desired information such as, e.g., closed orbit or
Twiss parameters.

It should therefore be the goal to generalize this concept to non-linear dynamics.
The computation of a reliable OTM and the analysis of its  properties
will provide all relevant information.

\subsubsection{Basic concept}
Given that the non-linear maps can be rather complex objects, the analysis of the
OTM should be separated from the calculation of the map itself.

Using the coordinate~vector ${\vec{z}} = (x, x' = {\partial x}/{\partial s}, y, y' = {\partial y}/{\partial s
})$,
the map ${{{\cal{M}}}}$ transforms the
coordinates $\vec{z_{1}}(s_{1})$ through a magnet {{$M$}} at position ${{s_{1}}}$ to
new coordinates $\vec{z_{2}}(s_{2})$ at position ${{s_{2}}}$:
\begin{eqnarray}
{{\vec{z_{2}}(s_{2})}} =
\left( \begin{array}{c}
x  \\
x' \\
y  \\
y' \\
\end{array}\right)_{{{s_{2}}}}
=
{{\cal{M}}}
 \circ
\left( \begin{array}{c}
x  \\
x' \\
y  \\
y' \\
\end{array}\right)_{{{s_{1}}}}
= {{\cal{M}}}
 \circ
{{\vec{z_{1}}(s_{1})}}.
\end{eqnarray}
The subsequent analysis of the map should reveal the quantities of interest.

\section{Linear maps}
First I shall introduce the simplest form of maps, i.e. linear maps, and generalize afterwards.

\subsection{Examples for linear maps}
As a first example, I consider a drift space of length {{$L$}}.

The formal description using the map can be written
\begin{eqnarray}
\left( \begin{array}{c}
x  \\
x' \\
\end{array}\right)_{s_{2}}
=
{{
\left( \begin{array}{cc}
1 &L \\
0 &1              \\
\end{array}\right)
}}
 \circ
\left( \begin{array}{c}
x  \\
x' \\
\end{array}\right)_{s_{1}}.
\end{eqnarray}
As a second example, I show the map (matrix) for a focusing quadrupole of length $L$ and strength $k$:
\begin{eqnarray}
\left( \begin{array}{c}
x  \\
x' \\
\end{array}\right)_{s_{2}}
=
{{
\left( \begin{array}{cc}
\cos{L\cdot \sqrt{k}} &\frac{1}{\sqrt{k}}\cdot \sin{L\cdot \sqrt{k}} \\
- \sqrt{k}\cdot \sin{L\cdot \sqrt{k}} &\cos{L\cdot \sqrt{k}} \\
\end{array}\right)
}}
 \circ
\left( \begin{array}{c}
x  \\
x' \\
\end{array}\right)_{s_{1}}.
\label{eq:001}
\end{eqnarray}

The map for a quadrupole with short length $L$ (i.e. $1 \gg L\cdot k$) can be approximated by
\begin{eqnarray}
\left( \begin{array}{c}
x  \\
x' \\
\end{array}\right)_{s_{2}}
=
{{
\left( \begin{array}{cc}
1 &0 \\
- k\cdot L(= \frac{1}{f}) &1              \\
\end{array}\right)
}}
 \circ
\left( \begin{array}{c}
x  \\
x' \\
\end{array}\right)_{s_{1}}.
\label{eq:002}
\end{eqnarray}
These are {$\cal{M}$}aps, and describe the movement within an element.

\subsection{OTMs}
Starting from a position {{$s_{0}$}} and applying all maps (for $N$ elements) in sequence around a ring with circumference $C$, we get the OTM for the position {{$s_{0}$}} (for one dimension only):
\begin{eqnarray}
\left( \begin{array}{c}
x  \\
x' \\
\end{array}\right)_{{{s_{0} + C}}}
=
{{\cal{M}}_{1}}
 \circ
{{\cal{M}}_{2}}
 \circ
\cdots
 \circ
{{\cal{M}}_{N}}
 \circ
\left( \begin{array}{c}
x  \\
x' \\
\end{array}\right)_{{{s_{0}}}}
\end{eqnarray}
\begin{eqnarray}
{{\Longrightarrow}}~~~
\left( \begin{array}{c}
x  \\
x' \\
\end{array}\right)_{{{s_{0} + C}}}
=
{{\cal{M}}_{\rm ring}(s_{0})}
 \circ
\left( \begin{array}{c}
x  \\
x' \\
\end{array}\right)_{{{s_{0}}}}.
\label{eq:01}
\end{eqnarray}

After this concatenation, we have obtained a OTM for the whole ring (\ref{eq:01}).
In the simplest case all maps are matrices and the concatenation is just the
multiplication of linear matrices.
Having obtained this OTM, its analysis will give all relevant information; there is no need for any assumptions.
To solve Hill's equation using the standard ansatz~\cite{bib:holzer}, we need several
assumptions such as existence of closed orbit and confinement or periodicity.
Having derived the map, we have to extract the information needed.
To introduce this topic, we first treat a linear OTM, i.e.\ a matrix.

\subsection{Linear normal forms}
The key to the analysis is that maps (here matrices) can be transformed into (Jordan) normal forms. The original maps and the normal forms are equivalent, but the normal form and the required transformation automatically provide the data we need, such as stability, closed orbit, optical parameters ($Q, Q'$, Twiss function, etc.), invariants of the motion, resonance analysis, and so on.

\subsubsection{Linear normal-form transformation}
The basic idea is to make a transformation to get a simpler form for the map.

Assuming that the map {{${\cal{M}}_{12}$}} propagates the variables from location $1$ to location $2$, we try to find transformations {{${\cal{A}}_{1}, {\cal{A}}_{2}$}} such that
\begin{eqnarray}
{\cal{A}}_{1}{{{\cal{M}}_{12}}}{\cal{A}}_{2}^{-1} = {{{\cal{R}}_{12}}}.
\end{eqnarray}
The map {{${\cal{R}}_{12}$}} is a `Jordan normal form' (or at least a very simplified form of the map). For example, {{${\cal{R}}_{12}$}} becomes a pure rotation, which is a simple and typical normal form. The map {{${\cal{R}}_{12}$}} describes the same dynamics as {{${\cal{M}}_{12}$}}, but all coordinates are transformed by ${\cal{A}}_{1}$ and ${\cal{A}}_{2}$. The transformations {{${\cal{A}}_{1}, {\cal{A}}_{2}$}} `analyse' the motion.

\begin{figure}
\centering\includegraphics[height= 2.50cm,width=  6.50cm]{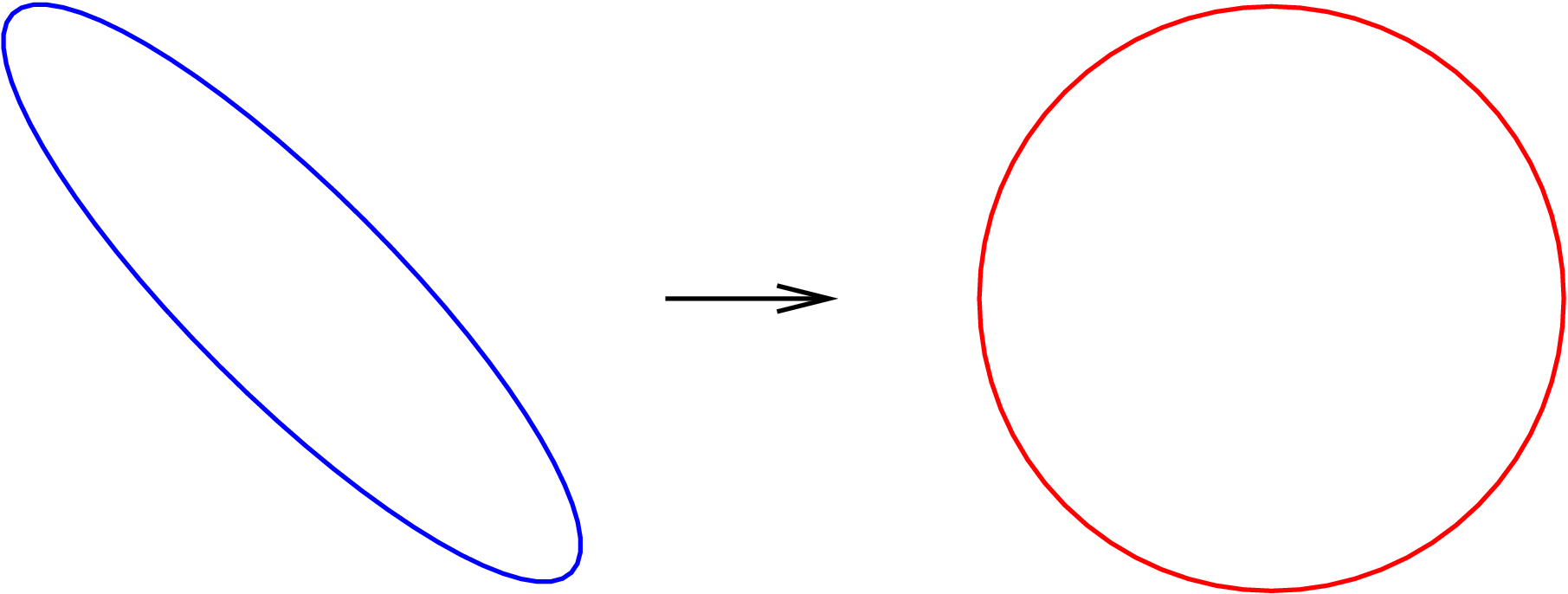}
\caption{Normal-form transform for a linear OTM (schematic)}
\label{fig:fig1}
\end{figure}

We assume that the OTM (here a matrix) ${\cal{M}}(s)$ at the position $s$ is
(see e.g.\ the article on transverse dynamics~\cite{bib:holzer})
\begin{eqnarray}
{\cal{M}}(s) =
\left( \begin{array}{cc}
\cos(\Delta\mu) + \alpha(s) \sin(\Delta \mu) &\beta(s) \sin(\Delta\mu)  \\
-\gamma(s) \sin(\Delta\mu) &\cos(\Delta\mu) - \alpha(s) \sin(\Delta \mu)   \\
\end{array}\right).
\end{eqnarray}
This matrix describes the motion on a phase-space ellipse. We can rewrite ${\cal{M}}$ such that one part ${\cal{R}}$ becomes a pure rotation (a circle), i.e.\
\begin{eqnarray}
{\cal{A}}{\cal{R}}{\cal{A}}^{-1} = {\cal{M}}.
\end{eqnarray}
Remembering lectures on linear algebra, this transformation is done by the
evaluation of the eigenvectors and eigenvalues.

Starting from
\begin{eqnarray}
 {{M}} = {\cal{A}} \circ {{{\cal{R}}(\Delta\mu)}} \circ {\cal{A}}^{-1}
\quad {\mathrm{or}} \quad {{{\cal{R}}(\Delta\mu)}} = {\cal{A}}^{-1} \circ {{M}} \circ {\cal{A}},
\end{eqnarray}
we get for the transformation (${\cal{A}}$) and the simple form (${\cal{R}}$)
\begin{eqnarray}
{\cal{A}} =
\left( \begin{array}{cc}
\sqrt{\beta(s)}  &0 \\
-\frac{\alpha}{\sqrt{\beta}} &\frac{1}{\sqrt{\beta(s)}}  \\
\end{array}\right)
\quad {\mathrm{and}} \quad {\cal{R}} =
\left( \begin{array}{cc}
\cos(\Delta\mu) &\sin(\Delta\mu)  \\
-\sin(\Delta\mu) &\cos(\Delta\mu)    \\
\end{array}\right).
\end{eqnarray}
This is just the Courant--Snyder transformation to get $\beta, \alpha$, etc.\ and $\Delta \mu$ is the tune. That is, the Courant--Snyder analysis is just a {{normal-form transform}} of the linear one-turn matrix. This transformation works in more than one dimension using the same formalism. The phase-space ellipse is transformed into a circle, which is the representation of the normal form, i.e.\ the rotation shown in~\Fref{fig:fig1}.

\begin{figure}[hb]
\begin{center}
\includegraphics[height= 3.20cm,width=  6.80cm]{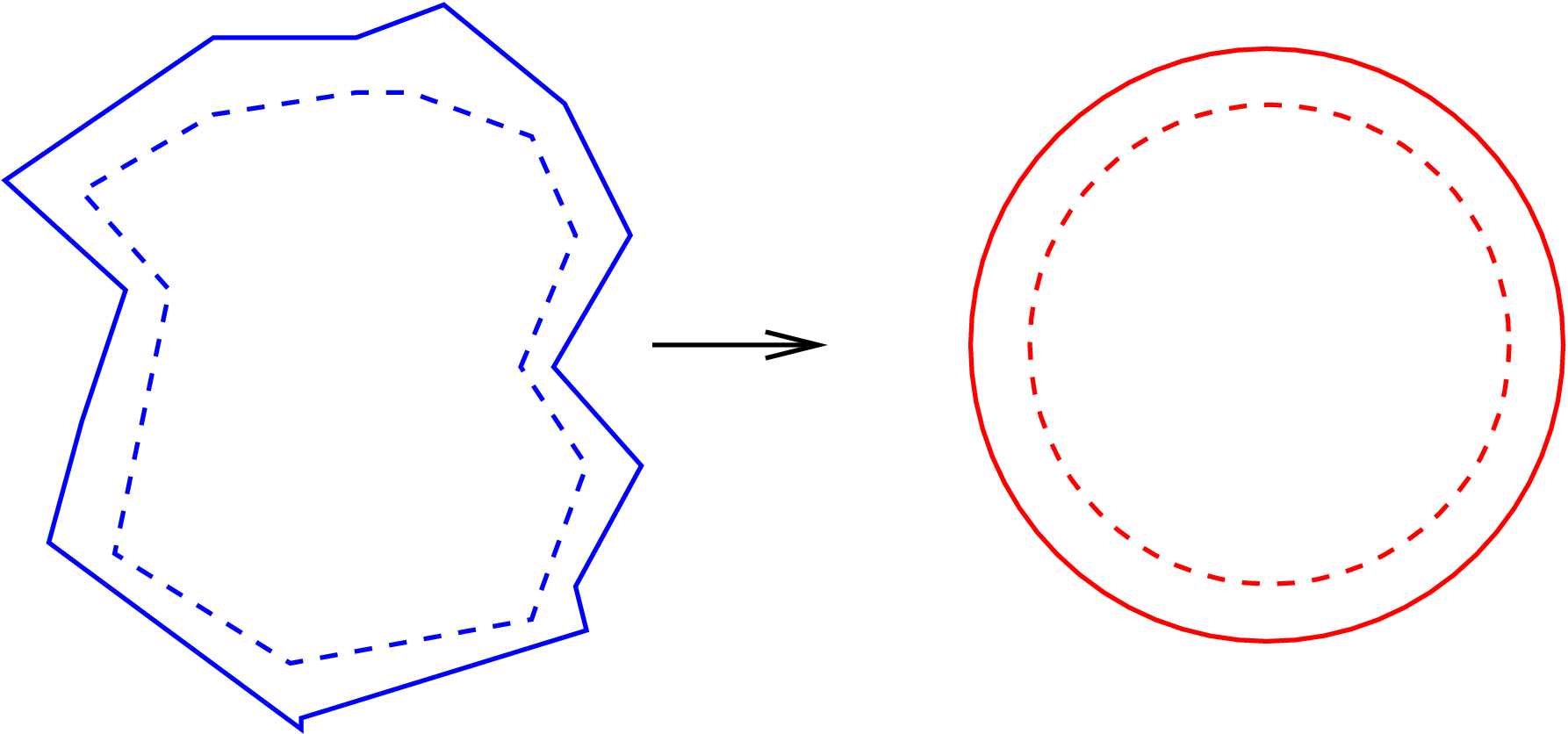}
\end{center}
\caption{Normal-form transform for a non-linear OTM (schematic)}
\label{fig:fig2}
\end{figure}

Compared with the Courant--Snyder analysis, we have the following properties of this method.
\begin{itemize}
\item[i)] The rotation angle (tune for the OTM) $\Delta \mu$ is contained in the {\textit{normalized map}}, and we have stability for real values of this phase advance $\Delta \mu$.
\item[ii)] Optical functions $(\beta, \alpha$, etc.) are in the {\textit{normalizing map}}.
\item[iii)] No need to make any assumptions, ansatz or approximation.
\end{itemize}
In the case of a more complicated OTM (e.g. \Fref{fig:fig2}), the transformation may be
more complicated and we have to expect that the rotation angle depends on the amplitude.
This is the case for non-linear OTMs, which will be treated in a later section.
It has already been mentioned here to introduce the normal-form transformation as a general
concept.

\subsubsection{Normalized variables}
Please note that
\begin{eqnarray}
\left( \begin{array}{c}
x_{n}  \\
x'_{n} \\
\end{array}\right)
=
{{\cal{A}}^{-1}}
 \circ
\left( \begin{array}{c}
x  \\
x' \\
\end{array}\right)
\end{eqnarray}
in this simple case is just a variable transformation to new, normalized variables.

\subsubsection{Action-angle variables}
It is common practice to define action-angle variables.
Once the particles `travel' on a circle, we can introduce new variables for simplification
and easier interpretation of the results.
\begin{itemize}
\item[i)] Radius (say: $\sqrt{2 J}$, with $J = \frac{x^{2}_{n} + x'^{2}_{n}}{2}$) is constant (invariant of motion): {\bf{action ${{J}}$}}.
\item[ii)] Phase advances by constant amount: {\bf{angle ${{\Psi}}$}}.
\end{itemize}
The invariant of motion $J$ should be the same at all locations in the ring and is separate from the angle $\psi$. For the propagation of coordinates around the ring, both are relevant. This separation of the invariant from the angle again becomes important in a later section.

\subsubsection{Coupling}
The formalism can easily be used to study the case of coupling between the two transverse planes, with the map (in two dimensions)
\begin{eqnarray}
T =
\left( \begin{array}{cc}
M     & n \\
m     &N \\
\end{array}\right),
\end{eqnarray}
where $M, m, N, n$ are $2 \times 2$ matrices. In the case of coupling, $m \neq 0, n \neq 0$.

We can try to rewrite this using the normal-form algorithms, as before:
\begin{eqnarray}
T =
\left( \begin{array}{cc}
M     & n \\
m     &N \\
\end{array}\right)
= V {{R}} V^{-1},
\end{eqnarray}
with
\begin{eqnarray}
{{R}} =
\left( \begin{array}{cc}
A     & 0 \\
0     &B \\
\end{array}\right)
\quad {\mathrm{and}} \quad
V =
\left( \begin{array}{cc}
\gamma I     & C \\
-C^{t} &\gamma I \\
\end{array}\right).
\end{eqnarray}
What have we obtained?
The matrix {{$R$}} is our simple rotation in two planes and we have in addition:
\begin{itemize}
\item[i)] $A$ and $B$ are the one-turn matrices for the `normal modes';
\item[ii)] the matrix $V$ transforms from the coordinates $(x, x', y, y')$ into the `normal-mode' coordinates $(w, w',v, v')$ via the expression
$(x, x', y, y') = V (w, w',v, v') $;
\item[iii)] the matrix $C$ contains the `coupling coefficients'.
\end{itemize}

\subsubsection{Multiple turns}
Normal forms are extremely useful when a map should be applied $k$ times (e.g.\ $k$ turns):
the brute-force method would be to multiply the OTM $k$ times, which can be
very elaborate and expensive in computing time.
Instead, we can first make a transformation into a normal form with a rotation matrix $R$,
which contains the phase advance per turn $\mu$.
For $k$ turns, we have to simply replace the phase advance $\mu$ by the phase advance for
$k$ turns, which is of course $k\cdot \mu$ (\ref{eq:02}).
\begin{eqnarray}
 M^{k}(x,x') = A R^{k}~A^{-1}(x,x') = A R^{k} (X, X').
\label{eq:02}
\end{eqnarray}
Transforming the rotation back into the usual form returns the
$k$-turn map.

\section{Non-linear maps}
In this section, I should like to introduce {\textit{non-linear maps}}.
I shall start with a linear map and than generalize.
The effect of a (short) quadrupole depends {{linearly}} on amplitude as
(rewritten from the matrix form)
\begin{eqnarray}
{\vec{z}(s_{2})} =
\left( \begin{array}{c}
x  \\
x' \\
y  \\
y' \\
\end{array}\right)_{s_{2}}
=
\left( \begin{array}{c}
x  \\
x' \\
y  \\
y' \\
\end{array}\right)_{s_{1}}
+
\left( \begin{array}{c}
0  \\
k_{1}\cdot {{x_{s_{1}}}}\\
0  \\
k_{1}\cdot {{y_{s_{1}}}}\\
\end{array}\right).
\label{eq:003}
\end{eqnarray}
Or, in the already known convention using the matrix ${\bf{M}}$,
${\vec{z}(s_{2})} = {\bf{M}} \cdot \vec{z}(s_{1}$).

\subsection{Symplecticity}
A key concept for the application of maps is a requirement called `symplecticity'.
In a ring, not all possible maps are allowed.
This requires for a matrix {${\cal{{{M}}}}$} that it fulfils the condition
\begin{eqnarray}
{{\cal{{M}}}^{\rm T} \cdot {{S}} \cdot \cal{{{M}}} = {{S}} }
\label{eq:03a}
\end{eqnarray}
with
\begin{eqnarray}
{{S}} =
\left( \begin{array}{cccc}
0  &1  &0  &0   \\
-1 &0  &0  &0   \\
0  &0  &0  &1   \\
0  &0  &-1 &0   \\
\end{array}\right).
\label{eq:03b}
\end{eqnarray}
The physical meaning of this requirement is that {${\cal{{{M}}}}$} is area preserving  and
\begin{eqnarray}
\lim_{n \rightarrow \infty} {\cal{{{M}}}}^{n} = {\mathrm{finite}} \quad
\Longrightarrow \quad {\mathrm{det}}~{\cal{{{M}}}} = 1.
\end{eqnarray}
It should be noted that ${\mathrm{det}}~{\cal{{{M}}}} = 1$ is required, but not sufficient, for ${\cal{{{M}}}}$ to be symplectic.

\subsection{Taylor maps}
The effect of a (thin) sextupole with strength $k_{2}$ is in the form of (\ref{eq:003})
\begin{eqnarray}
{\vec{z}(s_{2})} =
\left( \begin{array}{c}
x  \\
x' \\
y  \\
y' \\
\end{array}\right)_{s_{2}}
=
\left( \begin{array}{c}
x  \\
x' \\
y  \\
y' \\
\end{array}\right)_{s_{1}}
+
\left( \begin{array}{c}
0  \\
\frac{1}{2}k_{2}\cdot {{(x_{s_{1}}^{2} - y_{s_{1}}^{2})}}\\
0  \\
k_{2}\cdot {{(x_{s_{1}} \cdot y_{s_{1}})}}\\
\end{array}\right).
\label{eq:04}
\end{eqnarray}
We can write it now formally as
\begin{eqnarray}
{\vec{z}(s_{2})} = {\cal{M}} \circ \vec{z}(s_{1}).
\end{eqnarray}
Here {$\cal{M}$} is {{not}} a matrix but a non-linear map.
It cannot be written in linear matrix form.
We need something like (for the vector element $x$)
\begin{eqnarray}
\begin{array}{lll}
 \vec{z_{x}}(s_{2}) = x(s_{2})  &= & R_{11} \cdot {{x}^{  }}  + R_{12} \cdot {{x'}}  + R_{13} \cdot {{y}} +  \cdots.\\
 & &+ T_{111} \cdot {{x^{2}}} + T_{112} \cdot {{x x'}}  T_{122} \cdot {{x'^{2}}} +\\
 & &+ T_{113} \cdot {{x y}} + T_{114} \cdot {{x y'}} + \cdots\\
 & &+ U_{1111} \cdot {{x^{3}}} + U_{1112} \cdot {{x^{2} x'}} +  \cdots\\
\end{array}
\label{eq:05}
\end{eqnarray}
and the equivalent for other variables.
Please note that $x(s_{2})$ now depends also on the second transverse variable $y$.

\subsubsection{Forms of Taylor maps}
We can write (\ref{eq:05}) for $j = 1,\ldots,4$:
\begin{eqnarray}
z_{j}(s_{2}) = \sum_{k=1}^{4} {{R_{jk}}} z_{k}(s_{1}) +\sum_{k=1}^{4}\sum_{l=1}^{4} {{T_{jkl}}} z_{k}(s_{1})z_{l}(s_{1}).
\end{eqnarray}
If this map is truncated after the second order, we call it a second-order map and formally write it as
\begin{eqnarray}
{{\cal{A}}_{2}} = [{{R}},{{T}}].
\label{eq:06}
\end{eqnarray}
Higher orders can be defined as needed:
\begin{eqnarray}
{{\cal{A}}_{3}} = [R,T,{{U}}] \quad {{\Longrightarrow}} \quad [{{R}},{{T}}] + \sum_{k=1}^{4}\sum_{l=1}^{4}\sum_{m=1}^{4} {{U_{jklm}}} z_{k}(s_{1})z_{l}(s_{1})z_{m}(s_{1}),
\label{eq:07}
\end{eqnarray}
which is truncated after the third order.

As an example, the complete second-order map for a (thick) sextupole with length $L$ and strength {{$K$}} (in two dimensions) is
\begin{eqnarray}
\begin{array}{lll}
 x_{2}  &=  x_{1} + L x_{1}'  &- K \left(\frac{L^{2}}{4}(x_{1}^{2} - y_{1}^{2}) + \frac{L^{3}}{6}(x_{1}x_{1}' - y_{1}y_{1}') + \frac{L^{4}}{24}(x_{1}'^{2} - y_{1}'^{2}) \right),\\
 x_{2}'  &=  x_{1}'  &- K \left(\frac{L}{2}(x_{1}^{2} - y_{1}^{2}) + \frac{L^{2}}{2}(x_{1}x_{1}' - y_{1}y_{1}') + \frac{L^{3}}{6}(x_{1}'^{2} - y_{1}'^{2}) \right),\\
 y_{2}  &=  y_{1} + L y_{1}'  &+ K \left(\frac{L^{2}}{4}x_{1}y_{1} + \frac{L^{3}}{6}(x_{1}y_{1}' + y_{1}x_{1}') + \frac{L^{4}}{24}(x_{1}'y_{1}') \right),\\
 y_{2}'  &=  y_{1}'  &+ K \left(\frac{L}{2}x_{1}y_{1} + \frac{L^{2}}{2}(x_{1}y_{1}' + y_{1}x_{1}') + \frac{L^{3}}{6}(x_{1}'y_{1}') \right).\\
\end{array}
\label{eq:08}
\end{eqnarray}
The definition of {{$K$}} is not unique; it can differ by some factor, e.g.
\begin{eqnarray}
\left( \frac{\partial^{2} x}{\partial t^{2}} = k\cdot x^{2} \quad {\mathrm{versus}}
\quad \frac{\partial^{2} x}{\partial t^{2}} = \frac{k}{2}\cdot x^{2}\right).
\end{eqnarray}

\subsubsection{Symplecticity of Taylor maps}
We have argued that a vital requirement for maps is the symplecticity.
This must hold also for non-linear maps; however:
\begin{itemize}
\item[i)] truncated Taylor expansions are not matrices;
\item[ii)] it is the associated Jacobian matrix {{${\cal{J}}$}} which must fulfil the
symplecticity condition (see (\ref{eq:03a}) and (\ref{eq:03b})):
\begin{eqnarray}
{{\cal{J}}_{ik}} = \frac{\partial z^{i}_{2}}{\partial z^{k}_{1}} \quad \left({\mathrm{e.g.}}\ {{\cal{J}}_{xy}} = \frac{\partial z^{x}_{2}}{\partial z^{y}_{1}} \right),
\label{eq:09}
\end{eqnarray}
where ${\cal J}$ must fulfil ${\cal J}^{t} \cdot S \cdot {\cal J} = S$.
\end{itemize}

In general, ${{\cal{{{J}}}}_{ik}} \neq {\rm const }\rightarrow$
for a truncated Taylor map; this can be difficult to fulfil for all $z$.

As an example, we take the sextupole map (for simplicity in one dimension)
\cite{bib:wolski}
\begin{eqnarray}
\begin{array}{lll}
 x_{2}  &=  x_{1} + L x_{1}'  &- K \left(\frac{L^{2}}{4}x_{1}^{2} + \frac{L^{3}}{6}x_{1}x_{1}' + \frac{L^{4}}{24}x_{1}'^{2} + {\cal{O}}(3)\right),\\
 x_{2}'  &=  x_{1}'  &- K \left(\frac{L}{2}x_{1}^{2} + \frac{L^{2}}{2}x_{1}x_{1}' + \frac{L^{3}}{6}x_{1}'^{2} + {\cal{O}}(3)\right).\\
\end{array}
\label{eq:10}
\end{eqnarray}
Then we can compute
\begin{eqnarray}
{\cal{J}}^{\rm T}\cdot S\cdot {\cal{J}} =
\left( \begin{array}{cc}
0     &1 {{+ \Delta S}} \\
-1 {{- \Delta S}}    &0 \\
\end{array}\right)
 \neq S.
\label{eq:11}
\end{eqnarray}
We find that ${\cal{J}}$ is non-symplectic with the error
\begin{eqnarray}
 {{\Delta S}} = \frac{K}{72} L^{4} (L^{2}x'^{2} + 6L x x' + 6 x^{2}).
\label{eq:12}
\end{eqnarray}

\section{Thin elements}
From Eq.~(\ref{eq:12}), we may get a hint of how to attack this problem.
The error $\Delta S$ compared to the fully symplectic map depends on the fourth power
of the sextupole length $L^{4}$.
This can be extended to other elements as well and for short objects
the error may be negligible.
We can attempt to `make' all elements thin, i.e.\ short enough to have
a  negligible non-symplecticity.
For this purpose, we define
thin elements, i.e.\ an element with {\textit{no}} length,
because they avoid problems with
symplecticity of non-linear maps such as Taylor maps.
The error in equation (\ref{eq:12}) becomes zero.

\subsection{Concept of thin elements}
Real magnets have a finite length, i.e.\ they are thick magnets.
The field and length are used to compute their effect, i.e.\ the map; in the simplest case it becomes a matrix (\ref{eq:001}).
In passing, we have already mentioned the equivalent matrix for a quadrupole with zero length (\ref{eq:002}),
since it is often used in analytical calculations.
A severe consequence of thick elements is that they are not always linear elements (even not dipoles or quadrupoles).
For thick, non-linear magnets a closed solution for maps often does not exist.

\begin{figure}[t]
\centering\includegraphics[height= 2.20cm,width=  6.50cm]{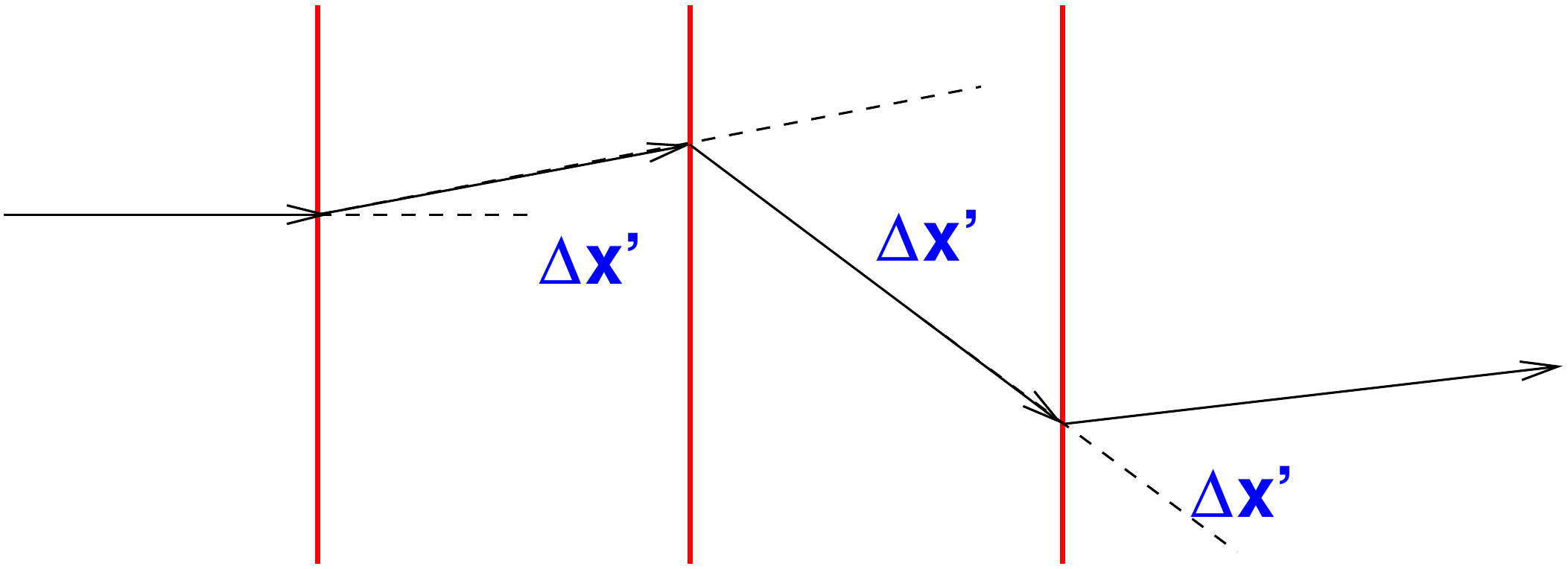}
\caption{Movement of a particle through thin elements}
\label{fig:fig3}
\end{figure}

The procedure we shall apply to get `thin' magnets is
to let the length go to zero, but keep the field integral finite (constant)
As consequences,
thin dipoles and quadrupoles are linear elements
and all thin magnets are much easier to use and automatically symplectic.
The movement of a particle through thin elements is shown schematically in \Fref{fig:fig3}.
The key is that there is no change of amplitudes $x$ and $y$ at the element; only
the momenta $x'$ and $y'$ receive an amplitude-dependent deflection {{(kick)}}
\begin{eqnarray}
x' \rightarrow x' + {{\Delta x'}} \quad {\mathrm{and}} \quad y' \rightarrow y' + {{\Delta y'}}.
\label{eq:13}
\end{eqnarray}
Modelling a magnet with a finite length by a thin magnet, we have to pay a price and
must ask under which conditions this is a good strategy.
We should expect that this approximation is valid when the finite length
is small or when the length does not matter.
The main questions to answer are as follows.
\begin{itemize}
\item[i)] What happens to the accuracy and the implications for the beam dynamics?
\item[ii)] What is the best implementation of thin magnets to minimize the side effects?
\end{itemize}
We should study these issues with some simple examples and generalize later.

\subsection{Accuracy of thin elements}
We start again with the exact matrix for a thick quadrupole with the strength $K$ and the length $L$.
\begin{eqnarray}
{\cal{M}}_{s \rightarrow s + L}=
{{
\left( \begin{array}{cc}
\cos{L\cdot \sqrt{K}}         &\frac{1}{\sqrt{K}}\cdot \sin{L\cdot \sqrt{K}} \\
-\sqrt{K}\cdot \sin{L\cdot \sqrt{K}}  &\cos{L\cdot \sqrt{K}} \\
\end{array}\right).
}}
\end{eqnarray}
This matrix describes the motion in a thick quadrupole exactly and we can estimate the
errors we make by our approximation.
Although this demonstration is done for a quadrupole, the final result is
valid for all linear and non-linear elements.

From the exact matrix
\begin{eqnarray}
{\cal{M}}_{s \rightarrow s + L}=
{{
\left( \begin{array}{cc}
\cos{L\cdot \sqrt{K}}         &\frac{1}{\sqrt{K}}\cdot \sin{L\cdot \sqrt{K}} \\
-\sqrt{K}\cdot \sin{L\cdot \sqrt{K}}  &\cos{L\cdot \sqrt{K}} \\
\end{array}\right),
}}
\end{eqnarray}
we make a Taylor expansion in `small' length {{$L$}}:
\begin{eqnarray}
{{L^{0}}} \cdot \left( \begin{array}{cc}
1 &0\\
0 &1\\
\end{array}\right)
+ {{L^{1}}} \cdot \left( \begin{array}{cc}
0 &1\\
-K &0\\
\end{array}\right)
+ {{L^{2}}} \cdot \left( \begin{array}{cc}
-\frac{K}{2} &0\\
0 &-\frac{K}{2}\\
\end{array}\right) + \cdots
\label{eq:14}
\end{eqnarray}
and keep up to the {\textit{first-order}} term in $L$:
\begin{eqnarray}
{\cal{M}}_{s \rightarrow s + L}=
{{L^{0}}} \cdot \left( \begin{array}{cc}
1 &0\\
0 &1\\
\end{array}\right)
+ {{L^{1}}} \cdot \left( \begin{array}{cc}
0 &1\\
-K &0\\
\end{array}\right),
\end{eqnarray}
\begin{eqnarray}
{\cal{M}}_{s \rightarrow s + L}=
{{
\left( \begin{array}{cc}
1           &L  \\
-K\cdot L  &1 \\
\end{array}\right) + {\cal{O}}(L^{2})
}}.
\label{eq:15}
\end{eqnarray}
We make two observations from the approximation (\ref{eq:15}).
\begin{itemize}
\item[i)]It is precise to first order ${\cal{O}}(L^{1})$ and the error is to second order ${\cal{O}}(L^{2})$.
\item[ii)]We have det~${\cal{M}} \neq 1$; therefore, it cannot be symplectic~(see \ref{eq:03a} and \ref{eq:03b}).
\end{itemize}
To ensure the symplecticity, we need to modify the matrix.
We can try to add a term in the matrix:
\begin{eqnarray}
{\cal{M}}_{s \rightarrow s + L}=
{{
\left( \begin{array}{cc}
1           &L  \\
-K\cdot L  &1 \\
\end{array}\right) + {{{\cal{O}}(L^{2})}}
}}
\end{eqnarray}
and we obtain
\begin{eqnarray}
{\cal{M}}_{s \rightarrow s + L}=
{{
\left( \begin{array}{cc}
1           &L  \\
-K\cdot L  &1 {{-K L^{2}}} \\
\end{array}\right)
}}.
\label{eq:16}
\end{eqnarray}
The matrix (\ref{eq:16}) is still precise to first order ${\cal{O}}(L^{1})$
but `symplectified' by adding the term {{$-K L^{2}$}}.
The added term is inaccurate to second order (because of $L^{2}$) and therefore the
order of accuracy is not changed by this modification.
This procedure restores the symplecticity, but does not
change the level of accuracy.

For the next step, we keep up to the {\textit{second order}} in $L$ from the Taylor expansion (\ref{eq:14}):
\begin{eqnarray}
{\cal{M}}_{s \rightarrow s + L}=
{{
\left( \begin{array}{cc}
1 - \frac{1}{2}K L^{2}          &L  \\
-K\cdot L  &1 - \frac{1}{2}KL^{2} \\
\end{array}\right) + {{{\cal{O}}(L^{3})}}
}}.
\label{eq:17}
\end{eqnarray}
It is precise to second order ${\cal{O}}(L^{2})$ and therefore
more accurate than (\ref{eq:16}), but again not symplectic.

We proceed as before and the symplectification now looks like
\begin{eqnarray}
{\cal{M}}_{s \rightarrow s + L}=
{{
\left( \begin{array}{cc}
1 - \frac{1}{2}KL^{2}          &L {{- \frac{1}{4}KL^{3}}} \\
-K\cdot L  &1 -\frac{1}{2}KL^{2} \\
\end{array}\right) + {{{\cal{O}}(L^{3})}}
}}.
\label{eq:18}
\end{eqnarray}
The accuracy is maintained to second order ${\cal{O}}(L^{2})$ and it is now fully symplectic.

\subsubsection{Physical meaning of the symplectification}
\begin{figure}[t]
\centering\includegraphics[height= 2.20cm,width=  6.00cm]{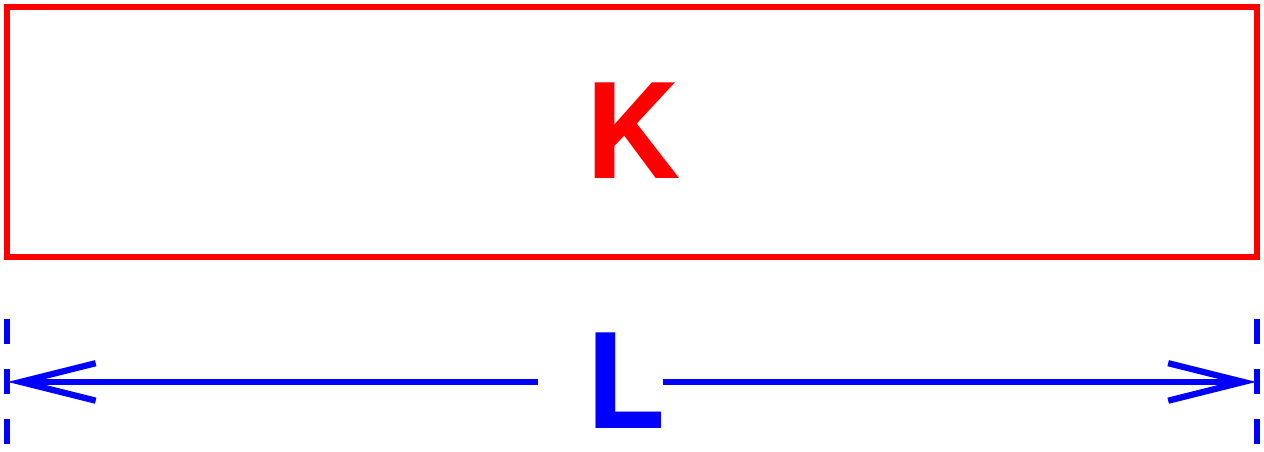}
\caption{Schematic picture of a quadrupole with length $L$ and strength $K$ (convention)}
\label{fig:fig4}
\end{figure}

\begin{figure}[t]
\centering\includegraphics[height= 2.20cm,width=  8.20cm]{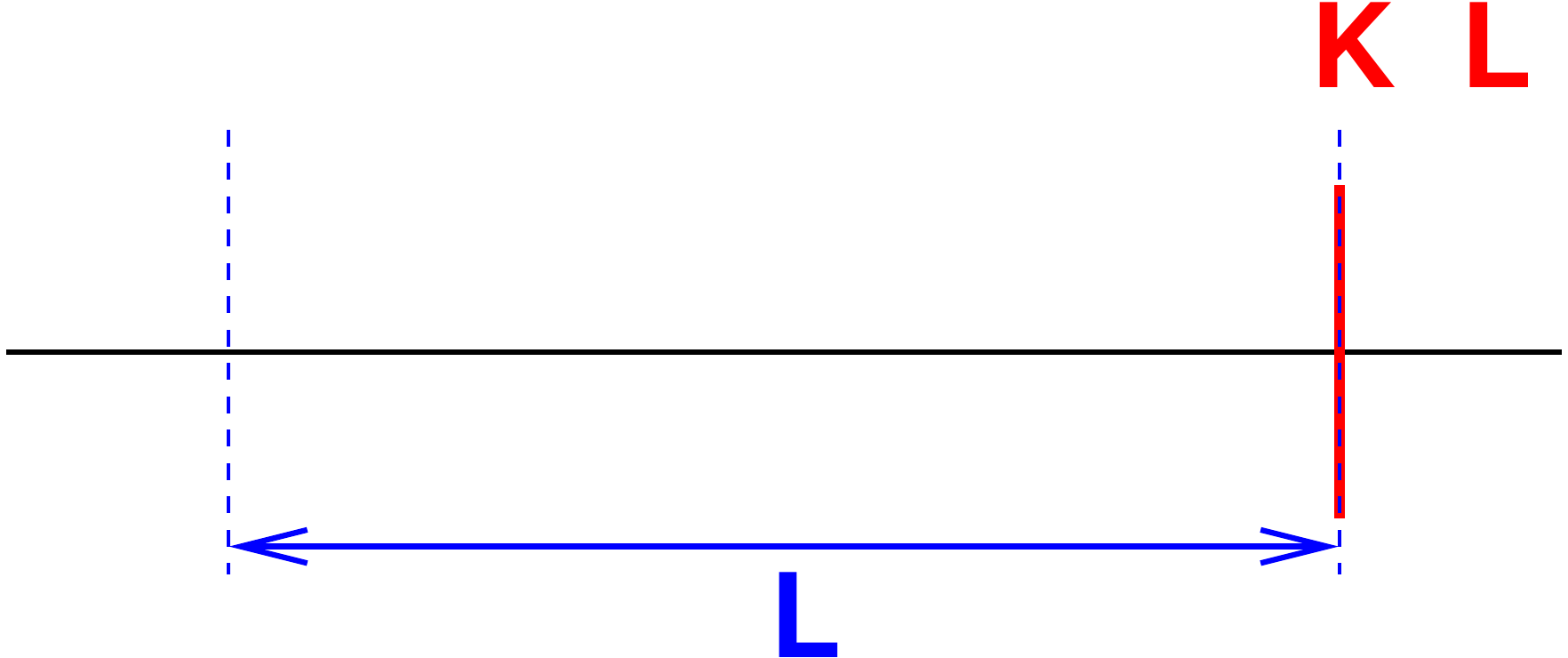}
\caption{Schematic picture of a quadrupole with length $L$ and a kick with strength $K$ at the end of the element}
\label{fig:fig6}
\end{figure}

In principle, one could go on with such a scheme, but
what is a good strategy?
The main issues for using an implementation are accuracy and speed.
First, we have a look at the significance of the symplectification,
which looks rather arbitrary.
Assume a linear element (quadrupole) of length $L$ and strength $K$ as
illustrated in (\Fref{fig:fig4}).
A very simple way to apply a thin-lens kick is shown in (\Fref{fig:fig6}).
The quadrupole is treated as a drift length of length $L$ followed by a
thin-length kick with the strength $K\cdot L$.

We can compute the full matrix of the drift and the kick and obtain
\begin{eqnarray}
\left( \begin{array}{cc}
1           &0  \\
-K\cdot L  &1 \\
\end{array}\right)
\left( \begin{array}{cc}
1           &L  \\
0  &1 \\
\end{array}\right) =
\left( \begin{array}{cc}
1           &L  \\
-K\cdot L  &1 -KL^{2} \\
\end{array}\right).
\label{eq:19}
\end{eqnarray}
We find that this resembles the `symplectification' we applied to the truncated map of order ${\cal{O}}(L^{1})$.
Because of the symmetry, we would get the same result when we apply the kick before the drift.

\begin{figure}[t]
\centering\includegraphics[height= 2.20cm,width=  8.20cm]{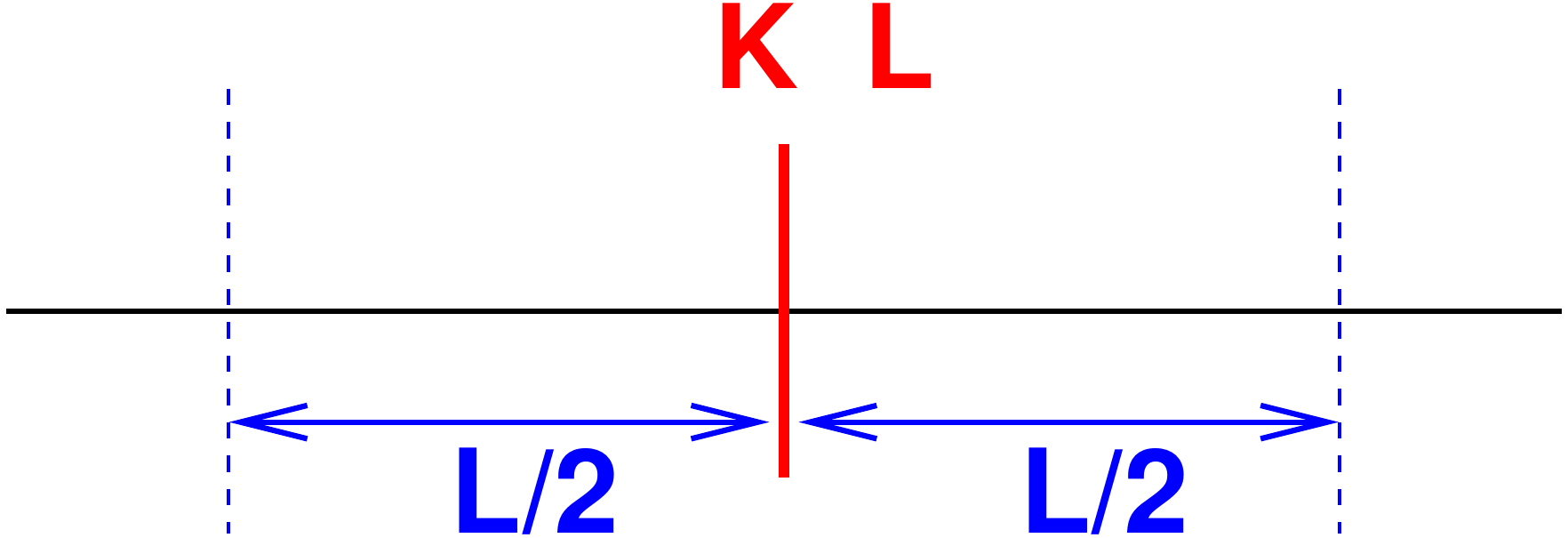}
\caption{Schematic picture of a quadrupole with length $L$ and a kick with strength $K$ in the centre of the element}
\label{fig:fig7}
\end{figure}

Another option is to apply the thin-lens kick in the centre of the element, as shown in \Fref{fig:fig7}.
This time we apply the kick in the centre of the element, preceded and followed by drift spaces of the length
$L/2$.

The multiplication of the three matrices leads us to
\begin{eqnarray}
{\cal{M}}_{s \rightarrow s + L}=
{{
\left( \begin{array}{cc}
1 - \frac{1}{2}KL^{2}          &L {{- \frac{1}{4}KL^{3}}} \\
-K\cdot L  &1 -\frac{1}{2}KL^{2} \\
\end{array}\right) + {{{\cal{O}}(L^{3})}}
}}.
\label{eq:20}
\end{eqnarray}
This is equivalent to the truncated and symplectified map of order ${\cal{O}}(L^{2})$.

We can summarize the two options as follows.
\begin{itemize}
\item[i)]One kick at the end or entry: error (inaccuracy) is of first order ${\cal{O}}(L^{1})$.
\item[ii)]One kick in the centre: error (inaccuracy) is of second order ${\cal{O}}(L^{2})$.
\end{itemize}
We find that it is very relevant how to apply thin lenses
and the aim should be to be precise and fast (and simple to implement, for example in
a computer program).

\subsection{Symplectic integration}
We introduce now the concept of symplectic integration and start with simple examples
and generalize the concept afterwards.
Different integration methods exist which yield symplectic maps and we shall discuss
some of them which are relevant for our discussion of applications to accelerators.

\subsubsection{Can we do better than in the previous examples?}
\begin{figure}[t]
\centering\includegraphics[height= 3.20cm,width= 10.20cm]{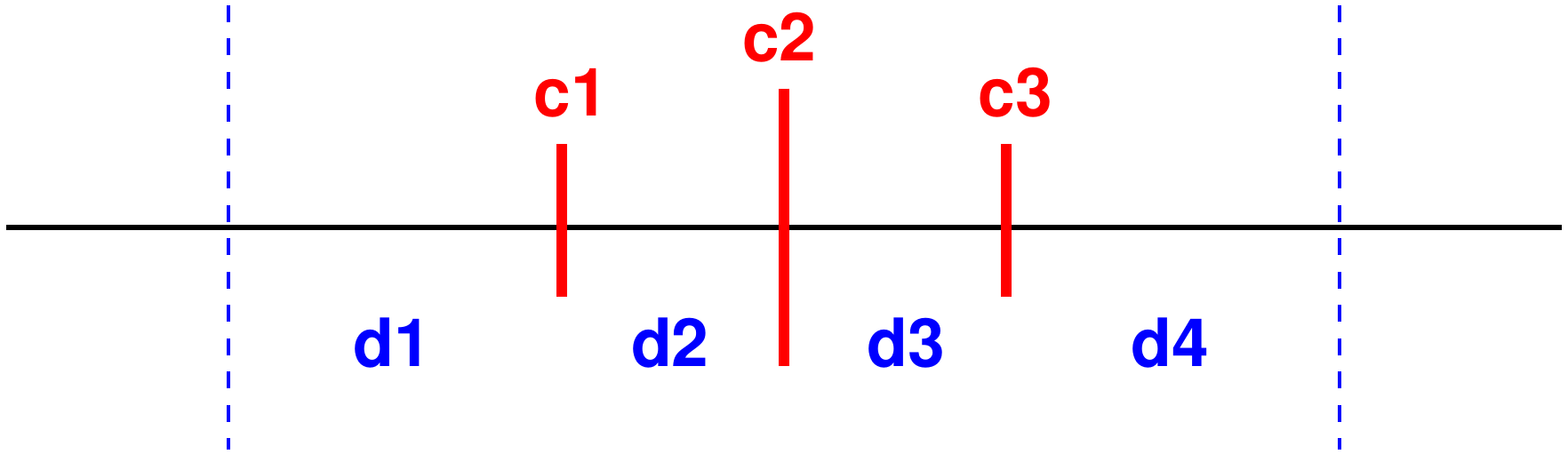}
\caption{Symplectic integrator with three thin lenses (schematic)}
\label{fig:fig8}
\end{figure}

We can try a model with three kicks~\cite{bib:ef02}, as shown in \Fref{fig:fig8}.
We assume that the kicks (c1, c2, c3) and drift spaces (d1, d2, d3, d4) are variable.
We can look for an optimization of these variables to get
the best accuracy (i.e.\ smallest deviation from the exact solution).
Since we have only thin lenses, the symplecticity is automatically ensured.

\begin{figure}[t]
\centering\includegraphics[height= 4.50cm,width= 10.20cm]{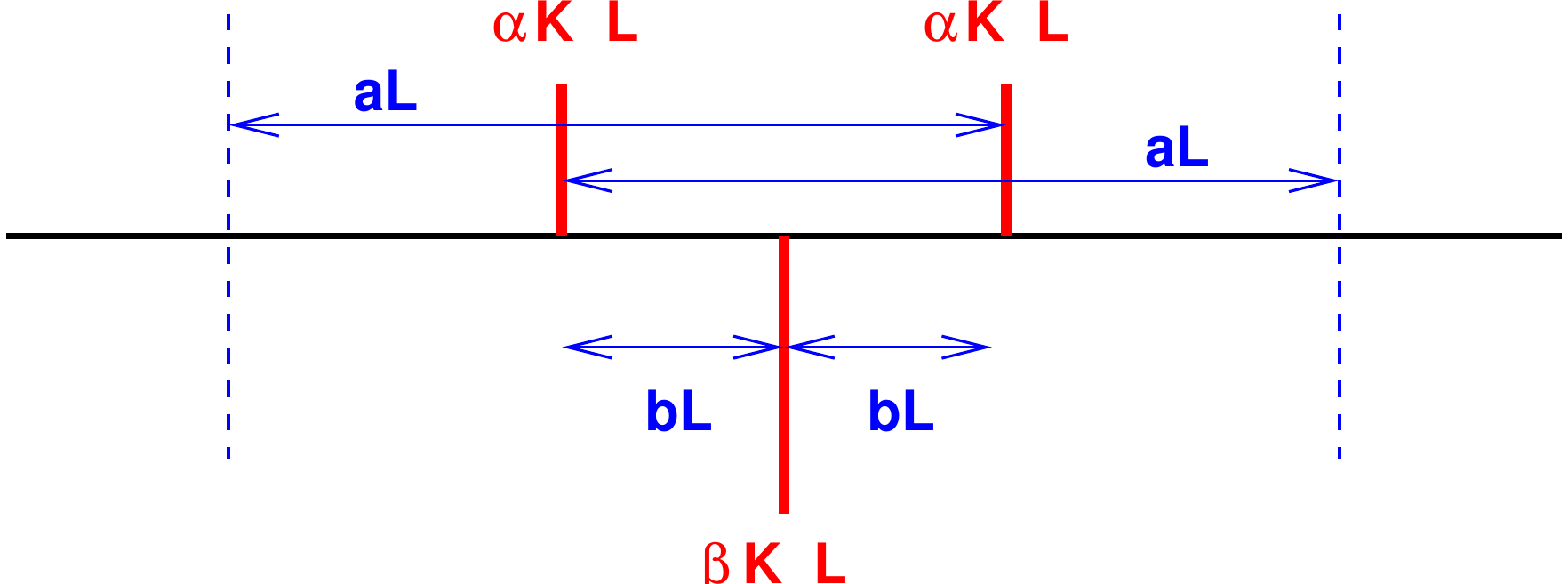}
\caption{Symplectic integrator with three thin lenses (schematic)}
\label{fig:fig9}
\end{figure}

The results of the optimization of the strengths and the drift spaces between thin kicks are shown in \Fref{fig:fig9}. The optimization of the free parameters gives the values (see \Fref{fig:fig9}): $a \approx 0.6756$, $b \approx -0.1756$,
$\alpha \approx 1.3512$, $\beta \approx -1.7024$.

It can easily be shown \cite{bib:ef02} that using this technique with three kicks, we obtain a ${\cal{O}}(4)$ integrator. We should expect that more kicks improve the accuracy and as an example a ${\cal{O}}(6)$ integrator would require nine kicks. This will be shown next.

\subsubsection{Yoshida formalism}
\begin{figure}[t]
\centering\includegraphics[height= 4.00cm,width= 10.20cm]{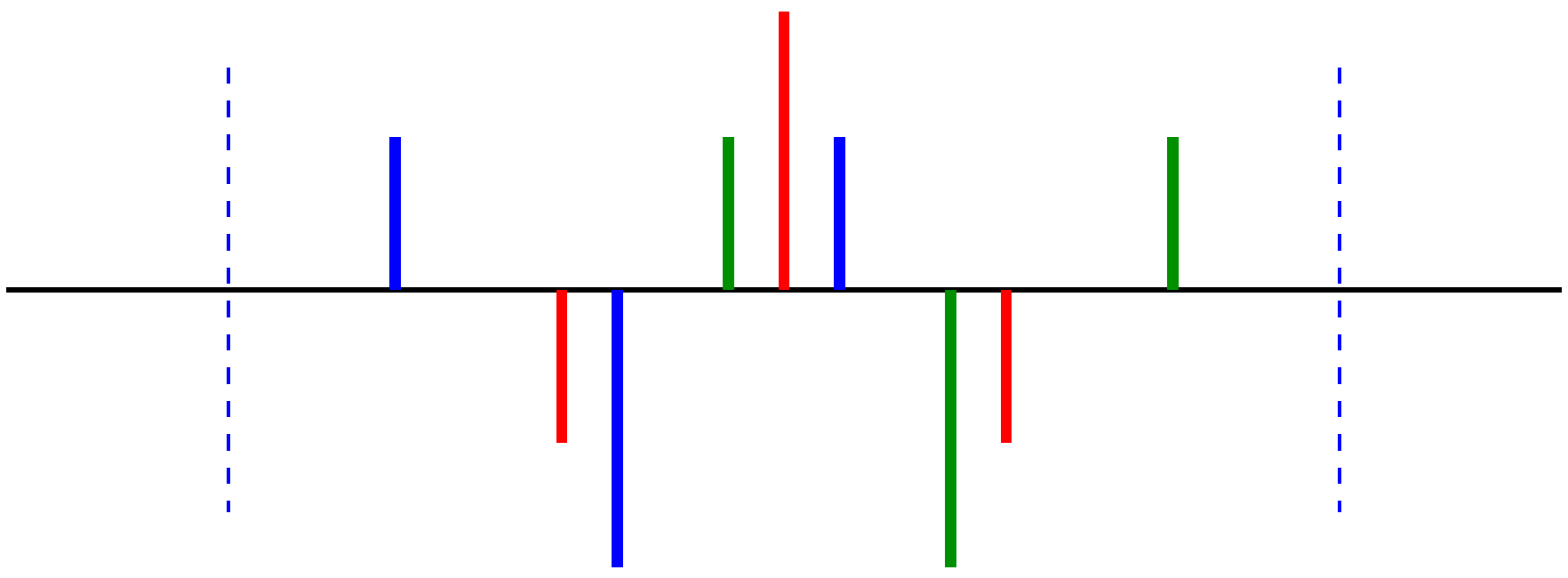}
\caption{Symplectic integrator with nine thin lenses (schematic)}
\label{fig:fig10}
\end{figure}

What we have done is a symplectic integration~\cite{bib:ef02, bib:yoshida}. From a lower order integration scheme (one kick), we can construct a higher order scheme. Formally (for the formulation of $S_{k}(t)$ see later), we can write: from a second-order scheme (one kick) $S_{2}(t)$ we construct a fourth-order scheme
(three kicks = $3 \times 1$  kick) like
\begin{eqnarray}
S_{4}(t) = S_{2}(x_{1}t) \circ S_{2}(x_{0}t) \circ S_{2}(x_{1}t)
\label{eq:44}
\end{eqnarray}
with the coefficients
\begin{eqnarray}
 x_{0} = \frac{-2^{1/3}}{2 - 2^{1/3}} \approx -1.7024,
 \quad x_{1} = \frac{1}{2 - 2^{1/3}} \approx 1.3512.
\label{eq:45}
\end{eqnarray}
Equation (\ref{eq:44}) should be understood symbolically in that we construct
a fourth-order integrator from three second-order integrators.
As a further example from a fourth order to a sixth order, we have to use
\begin{eqnarray}
S_{6}(t) = S_{4}(x_{1}t) \circ S_{4}(x_{0}t) \circ S_{4}(x_{1}t).
\end{eqnarray}
The interpretation of this construct is that we get three times fourth order with three kicks each; we have the nine-kick, sixth-order integrator mentioned earlier. This is shown schematically in \Fref{fig:fig10}. The three different colours correspond to the three fourth-order integrators, put together into a sixth-order integrator. The sixth-order integrator requires nine kicks and we have three interleaved fourth-order integrators. Going to higher order such as eighth order, we need three interleaved sixth-order integrators, altogether 27 thin lenses. The key is that this can be used as an iterative scheme to go to very high orders and can easily be implemented on a computer. More importantly, the coefficients in (\ref{eq:45}) are the same in this iterative procedure.

As a general formula, we can express this procedure as
\begin{eqnarray}
S_{k+2}(t) = S_{k}(x_{1}t) \circ S_{k}(x_{0}t) \circ S_{k}(x_{1}t).
\end{eqnarray}
In practice, this means that each kick in the $k$th-order integrator is replaced
by a $k$th-order integrator scaled accordingly to obtain a scheme of $(k+2)$th order.

We have used a linear map (quadrupole) to demonstrate the integration, but
it can be applied to other maps (solenoids, higher-order, non-linear maps).
It was shown~\cite{bib:ef02, bib:yoshida} that this is indeed possible and,
most importantly, we get the same integrators.
The proof and the systematic extension can be done in the form of Lie operators.
It can be proven that this scheme gives the theoretically best possible accuracy for
a given number of thin elements.

\subsubsection{Integration of non-linear elements}
We demonstrate now this formalism for {{non-linear}} elements.
Let us assume a general case:
\begin{eqnarray}
x'' &= &f(x).
\label{eq:41}
\end{eqnarray}
The disadvantage of a kick (\ref{eq:41}) is that usually a closed solution through the
element does not exist and an integration is necessary.

The advantage of non-linear elements is that they are usually thin (at least thinner than dipoles or quadrupoles). As examples, we give the lengths of typical LHC magnets.
\begin{itemize}
\item[i)]Dipoles: $\approx$ 14.3 m.
\item[ii)]Quadrupoles: $\approx$ 2--5 m.
\item[iii)]Sextupoles, octupoles: $\approx$ 0.30 m.
\end{itemize}
We can hope to obtain a good approximation for elements with small lengths. To get an estimate, we try our simplest ${\cal{O}}(2)$ thin-lens approximation first. Therefore, we represent the non-linear element as a thin-lens kick (\ref{eq:41}) with drift spaces of $L/2$ before and after the kick.

\subsubsection{Accuracy of thin lenses: our ${\cal{O}}(2)$ model}
Assuming the general case for a non-linear kick $\Delta x'$,
\begin{eqnarray}
x'' &= &f(x)\quad(= {{\Delta x'}}),
\label{eq:21}
\end{eqnarray}
we follow the three steps:
\begin{eqnarray}
\mathrm{Step~1.}\quad\left( \begin{array}{c}
x  \\
x' \\
\end{array}\right)_{s_{1} + L/2}
=
{{
\left( \begin{array}{cc}
1 &\frac{L}{2} \\
0 &1              \\
\end{array}\right)
}}
 \circ
\left( \begin{array}{c}
x  \\
x' \\
\end{array}\right)_{s_{1}};
\end{eqnarray}

\begin{eqnarray}
\mathrm{Step~2.}\quad\left( \begin{array}{c}
x  \\
x' \\
\end{array}\right)_{s_{1} + L/2}
=
{{
\left( \begin{array}{c}
x  \\
x' + {{\Delta x'}}\\
\end{array}\right)_{s_{1} + L/2}
}};
\end{eqnarray}

\begin{eqnarray}
\mathrm{Step~3.}\quad\left( \begin{array}{c}
x  \\
x' \\
\end{array}\right)_{s_{1} + L}
=
{{
\left( \begin{array}{cc}
1 &\frac{L}{2} \\
0 &1              \\
\end{array}\right)
}}
 \circ
\left( \begin{array}{c}
x  \\
x' \\
\end{array}\right)_{s_{1} + L/2}.
\end{eqnarray}

\begin{figure}[t]
\centering{\includegraphics*[width=50.1mm,height=25.0mm]{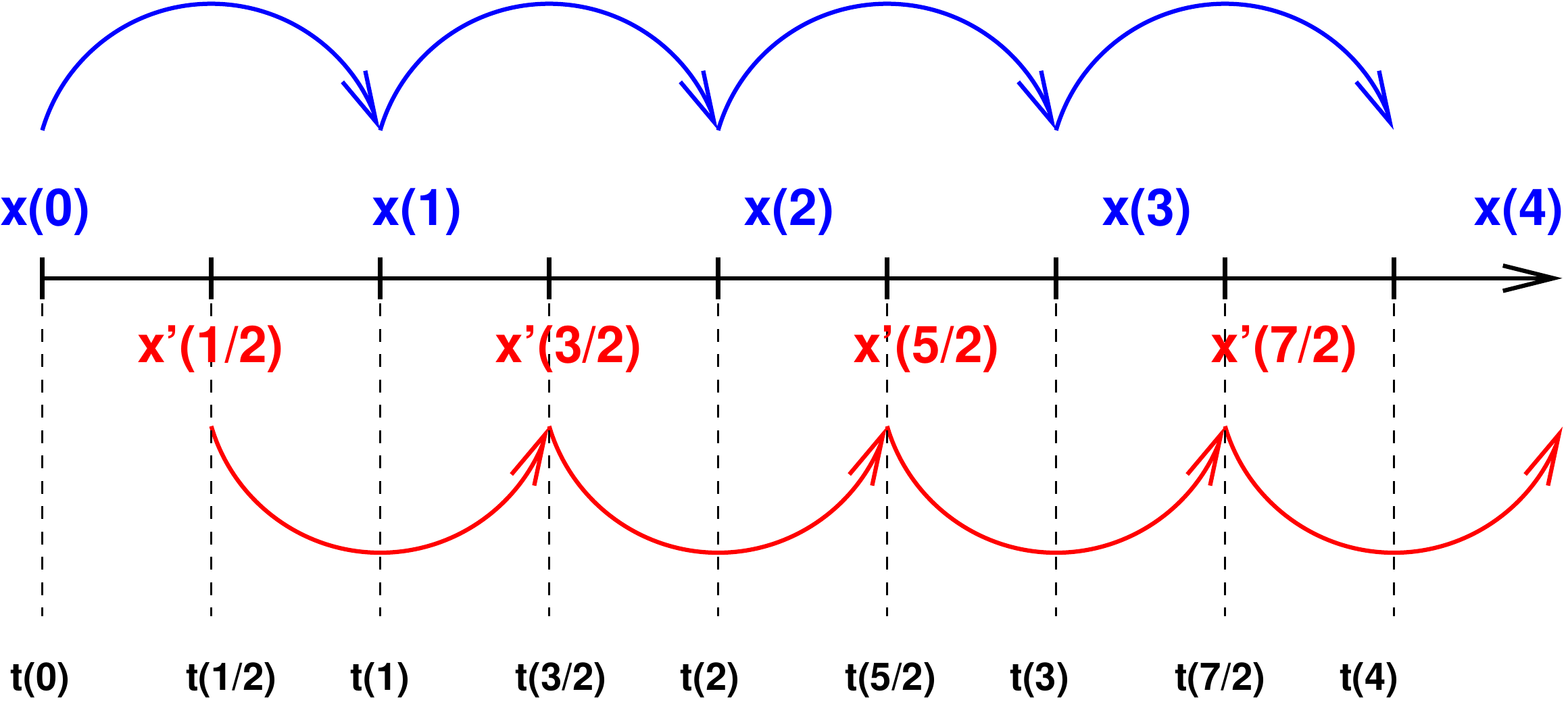}
\includegraphics*[width=34.0mm,height=30.0mm]{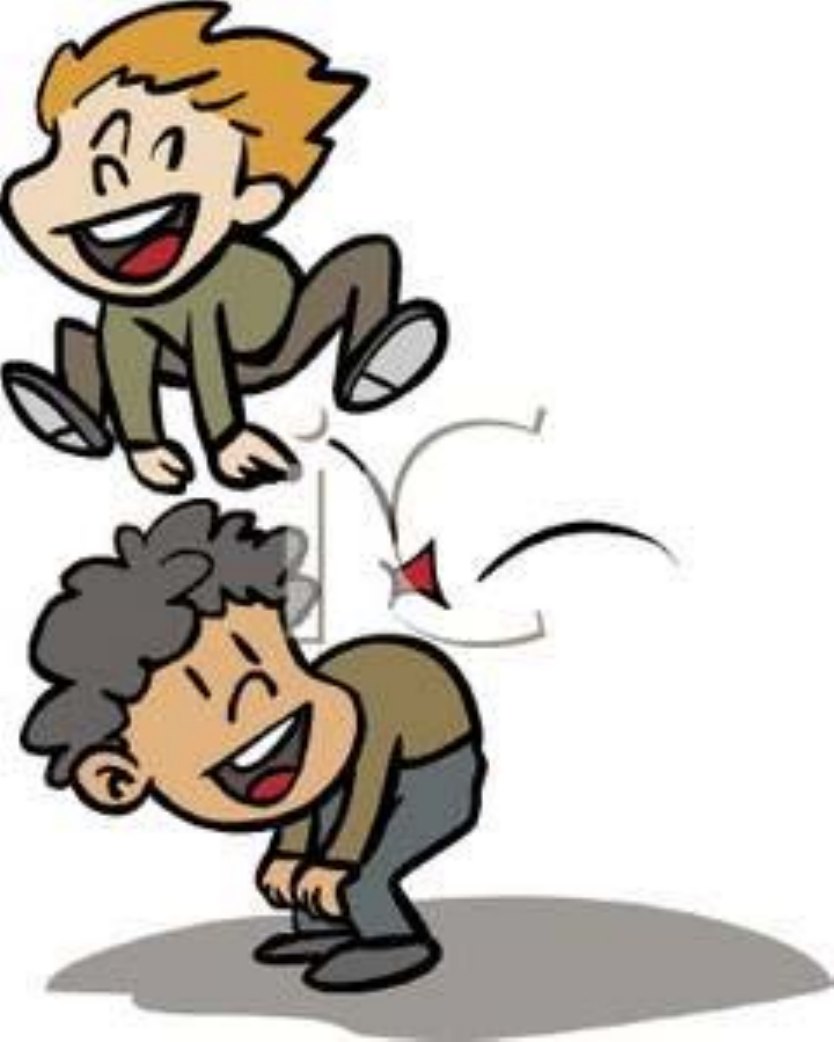}}
\caption{Schematic representation of a `leap-frog' integration scheme}
\label{fig:fig11}
\end{figure}

Using this thin-lens approximation (${\cal{O}}(2)$) gives us for the map
\begin{eqnarray}
x'(L) &\approx &x'_{0} + L f(x_{0} + \frac{L}{2}x'_{0}), \nonumber \\
x(L) &\approx  &x_{0}  + \frac{L}{2} (x'_{0} + x'(L)).
\label{eq:22}
\end{eqnarray}
Here we write $x_{0}$ and $x'_{0}$ for the coordinates and angles at the entry of the element. The result (\ref{eq:22}) corresponds to the well-known `leap-frog' algorithm/integration, schematically shown in~\Fref{fig:fig11}. It is symplectic and time reversible.

For any $x'' = f(x, x', t)$, we can solve it using
\begin{eqnarray}
x'_{i+3/2} &\approx &x'_{i+1/2} + f(x_{i+1}) \Delta t, \nonumber \\
x_{i+1} &\approx  &x_{i}  + x'_{i+1/2} \Delta t,
\label{eq:23}
\end{eqnarray}
which is what is known as a `leap-frog' integration.

\subsubsection{Accuracy of the `leap-frog' algorithm/integration}
The (exact) Taylor expansion gives
\begin{eqnarray}
x(L)&= &x_{0} + x'_{0}L + \frac{1}{2}f(x_{0})L^{2}
+ {{\frac{1}{6}}}x'_{0}f'(x_{0})L^{3} + \cdots;
\label{eq:24}
\end{eqnarray}
the (approximate) `leap-frog' algorithm (\ref{eq:23}) gives
\begin{eqnarray}
x(L)&= &x_{0} + x'_{0}L + \frac{1}{2}f(x_{0})L^{2} + {{\frac{1}{4}}}x'_{0}f'(x_{0})L^{3} +\cdots.
\label{eq:25}
\end{eqnarray}
The two methods differ from $L^{3}$ onwards and we have an integrator of order ${\cal{O}}(L^{3})$.
This had to be expected since we know from previous considerations that this type of
splitting gives such an accuracy.
For small $L$, this method is acceptable and symplectic.

As an application, we assume a (one-dimensional) sextupole with
(definition of $k$ not unique)
\begin{eqnarray}
x'' &= &k \cdot x^{2}  =  f(x);
\label{eq:26}
\end{eqnarray}
using the thin-lens approximation using (\ref{eq:22}) gives
\begin{eqnarray}
x(L) &=  &x_{0}  + x'_{0}L + \frac{1}{2}kx^{2}_{0}L^{2}
+ \frac{1}{2}kx_{0}x'_{0}L^{3} + \frac{1}{8}kx'^{2}_{0}L^{4}, \nonumber\\
x'(L) &= &x'_{0} + k x^{2}_{0}L + kx_{0}x'_{0}L^{2} + \frac{1}{4}kx'^{2}_{0}L^{3}.
\label{eq:27}
\end{eqnarray}
This is the map for a thick sextupole of length $L$ in thin-lens approximation, accurate to ${\cal{O}}(L^{2})$.

\section{Hamiltonian formalism}
In this section, we introduce the Hamiltonian formalism for
the classical motion of a particle~\cite{bib:gold01}.
We describe the motion of a system by a function $L$:
\begin{eqnarray}
L(q_{1},\ldots,q_{n}, \dot{q_{1}},\ldots,\dot{q_{n}}, t),
\end{eqnarray}
where
($q_{1},\ldots ,q_{n}$) are generalized coordinates and
($\dot{q_{1}},\ldots ,\dot{q_{n}}$) are generalized velocities.
For the introduction of the concepts, we use $q$ for the coordinate,
since it is commonly used in the literature~\cite{bib:gold01}.
The function $L$ defines the {{Lagrange function}} and
the integral
\begin{eqnarray}
I = \int L(q_{i},\dot{q_{i}},t) \, {\mathrm{d}}t
\end{eqnarray}
defines the action.
Without proof or derivation, we quote~\cite{bib:gold01}
\begin{eqnarray}
 L = T - V,
\end{eqnarray}
where $T$ is the kinetic energy and $V$ the potential energy.

\begin{figure}[t]
\begin{center}
{\includegraphics*[width= 62.2mm,height=55.2mm]{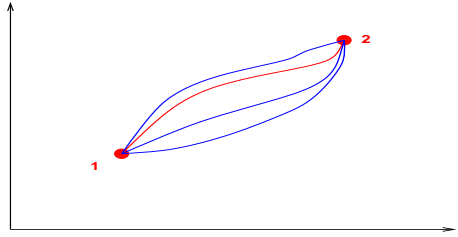}}
\end{center}
\caption{Schematic of the Hamiltonian principle. Different paths and path with minimum action}
\label{fig:fig12}
\end{figure}

The Hamiltonian principle states that a system moves such that the action $I$ becomes an extremum when different paths are possible. This is schematically shown in \Fref{fig:fig12}. The action becomes a minimum for the red path. Such concepts are known from other applications such as optics or quantum mechanics using Feynman path integrals.

Without proof, the action
\begin{eqnarray}
 I = \int_{1}^{2} L(q_{i},\dot{q_{i}},t) \, {\mathrm{d}}t = \mathrm{extremum}
\end{eqnarray}
becomes an extremum (minimum) when the Lagrange function fulfils the relation~\cite{bib:gold01}
\begin{eqnarray}
\frac{{\rm d}}{{\rm d}t}\frac{\partial L}{\partial \dot{q_{i}}} - \frac{\partial L}{\partial q_{i}} = 0,
\end{eqnarray}
which is usually called the Euler--Lagrange equation.

\subsection{From the Lagrangian to the Hamiltonian}
The Lagrangian $L(q_{1},\ldots ,q_{n}, \dot{q_{1}},\ldots ,\dot{q_{n}}, t)$ in generalized coordinates and velocities provides ($n$) second-order differential equations. We can try to get first-order differential equations (which are always solvable) and generalized momenta instead of velocities. This leads us to the Hamiltonian function and the Hamiltonian formalism. The main differences are as follows.
\begin{itemize}
\item[i)]Lagrangian:
\begin{itemize}
\item[a)]$n$ second-order equations;
\item[b)]$n$-dimensional {{coordinate space}}.
\end{itemize}
\item[ii)]Hamiltonian:
\begin{itemize}
\item[a)]$2n$ first-order equations;
\item[b)]$2n$-dimensional {{phase space}}.
\end{itemize}
\end{itemize}

Corresponding (so-called conjugate) momenta $p_{j}$ are
\begin{eqnarray}
p_{j} = \frac{\partial L}{\partial \dot{q_{j}}}.
\end{eqnarray}
Once we know what the canonical momenta {{$p_{i}$}} are,
the {{Hamiltonian}} is a transformation of the {{Lagrangian}}:
\begin{eqnarray}
H(q_{j},{{p_{j}}},t) = \sum_{i} {\dot{q_{i}}}{{p_{i}}} - L(q_{j},\dot{q_{j}},t).
\end{eqnarray}
Without proof or derivation, we quote~\cite{bib:gold01}
\begin{eqnarray}
 H = T + V,
\end{eqnarray}
where $T$ is the kinetic energy and $V$ the potential energy.

From the Hamiltonian principle, we obtain two first-order equations of motion:
\begin{eqnarray}
  \frac{\partial H}{\partial q_{j}} = -\dot{p_{j}} = -\frac{{\rm d} p_{j}}{{\rm d}t},
  \quad \frac{\partial H}{\partial p_{j}} = \dot{q_{j}} = \frac{{\rm d} q_{j}}{{\rm d}t}.
\label{eq:31}
\end{eqnarray}
The key difference to the Lagrangian is that the canonical momenta can be considered as independent
variables (contrary to velocities of the coordinates) and are on equal footing with the coordinates.
They form a separate set of variables and their time development is treated separately (\ref{eq:31}).

\subsection{Hamiltonian of electromagnetic elements in accelerators}
For the Hamiltonian of a (relativistic) particle in an electromagnetic field, we have (using from now on the more familiar variable $x$ as the coordinate)
\begin{eqnarray}
{\cal{H}}(\vec{x},\vec{p}, t) = c \sqrt{(\vec{p} - e \vec{A}(\vec{x}, t) )^{2}
+ m_{0}^{2}c^{2}} + e \Phi(\vec{x}, t),
\label{eq:32}
\end{eqnarray}
where $\vec{A}(\vec{x}, t)$,  $\Phi(\vec{x}, t)$ are the vector and scalar potentials.
Using canonical variables and the design path length $s$ as independent variable (bending in $x$-plane),
\begin{eqnarray}
{\cal{H}} =  -\left(1 + \frac{x}{\rho}\right) \cdot \sqrt{(1 + \delta)^{2} - p_{x}^{2} - p_{y}^{2}} +
\frac{x}{\rho} + \frac{x^{2}}{2\rho^{2}} - \frac{A_{s}(x, y)}{B_{0} \rho},
\end{eqnarray}
where $\delta = (p - p_{0})/p$ is the relative momentum deviation
and $A_{s}(x, y)$ the longitudinal component of the vector potential.

In general, the magnetic fields can be described with the multipole expansion
\begin{eqnarray}
B_{y} + {\rm i} B_{x} = \sum_{n=1} (b_{n} + {\rm i} a_{n}) (x + {\rm i} y)^{n-1}
\end{eqnarray}
and, since $\vec{B} = \nabla \times \vec{A}$ for the vector potential $\vec{A}$,
\begin{eqnarray}
A_{s}(x, y) = \sum_{n=1} \frac{1}{n} [(b_{n} + {\rm i} a_{n}) (x + {\rm i} y)^{n}].
\end{eqnarray}
Here $n = 1$ refers to dipolar fields (not always the case and different conventions exist).
For a large machine ($x \ll \rho$), we expand the root and sort the variables to
get
\begin{eqnarray}{\cal{H}} =
\overbrace{\frac{p_{x}^{2} + p_{y}^{2}}{2(1 + \delta)}}^{kinematic} -
\overbrace{\underbrace{\frac{x\delta}{\rho}}_{dispersive}
+ \underbrace{\frac{x^{2}}{2\rho^{2}}}_{focusing}}^{dipole} +
\overbrace{\frac{k_{1}}{2}(x^{2} - y^{2})}^{quadrupole} +
\overbrace{\frac{k_{2}}{6}(x^{3} - 3xy^{2})}^{sextupole}
\label{eq:33}
\end{eqnarray}
\begin{eqnarray}
\left(\mbox{using~MAD~convention:}\quad k_{n} =
\frac{1}{B\rho}\frac{\partial^{n}B_{y}}{\partial x^{n}} \right).
\end{eqnarray}
We can see the contribution to the different multipole fields.
The Hamiltonian describes {\textit{exactly}} the motion of a
particle {\textit{through a magnet}}.
This is our basis to extend the linear to a non-linear formalism.

But how do we apply it to an accelerator? A standard method used in the past was to derive a grandiose Hamiltonian for the whole ring. In that case one views the machine elements (magnets) as a localized fluctuation in the variable $s$, requiring a large set of Fourier modes. In this traditional treatment one tries to find a global description of the entire ring to derive the aspects of beam dynamics of interest. This goal is usually doomed to fail for complex systems, in particular with localized strong and non-linear objects such as beam--beam interactions in a collider. It is the humble opinion of the author that such an approach does not reach beyond (doubtful) pedagogical purposes.

Instead, I shall treat individual magnets as objects having an existence on their own.
The magnets are treated as $local$ objects where we can compute the map for the magnet.
In a global view it is not possible to have an idea of the effect of a single magnet
(e.g.\ a quadrupole) unless the Twiss parameters at the locations of the magnets are
known. This requires a priori knowledge of the OTM.

This `local' approach is clearly influenced by the computer simulation of accelerators where
conceptually all elements are treated as separate objects, which are described (and used)
independently of `global' concepts such as closed orbits, beta functions etc.
With this treatment the properties of the ring such as stability and optical
functions will arise naturally, as demonstrated in the simple introduction to
linear normal forms.
The description of machine elements can be easily achieved by defining the maps through
their individual Hamiltonians.
This allows naturally the treatment of non-linear elements and in connection to
simulations it translates itself into the `object-oriented' approach in computer
science.
Concepts built on this approach are described in detail in~\cite{bib:forest}.

In the next step we proceed to present the necessary tools.

\subsection{Poisson brackets}
We introduce a Poisson bracket for a differential operator~\cite{bib:gold01}:
\begin{eqnarray}
[f, g] = \sum_{i=1}^{n} \left( \frac{\partial f}{\partial x_{i}}\frac{\partial g}{\partial p_{i}} - \frac{\partial f}{\partial p_{i}}\frac{\partial g}{\partial x_{i}}\right).
\label{eq:34}
\end{eqnarray}
Here the variables $x_{i}, p_{i}$ are canonical variables; $f$ and $g$ are any functions of $x_{i}$ and $ p_{i}$.

We can now write (using the Hamiltonian $H$ for the function $g$ in (\ref{eq:34})
\begin{eqnarray}
f(x_{i}, p_{i}) = x_{i} \Rightarrow ~~[x_{i}, H] = \frac{\partial H}{\partial p_{i}} = \frac{{\rm d} x_{i}}{{\rm d}t},
\end{eqnarray}
\begin{eqnarray}
f(x_{i}, p_{i}) = p_{i} \Rightarrow ~~[p_{i}, H] = -\frac{\partial H}{\partial x_{i}} = \frac{{\rm d} p_{i}}{{\rm d}t}.
\end{eqnarray}
Therefore, Poisson brackets encode Hamiltons's equations~\cite{bib:gold01}.
In general, the Poisson bracket $[f, H]$ describes the time evolution of the system (i.e.\ the function $f$).
It is a special case of
\begin{eqnarray}
\frac{{\rm d}f}{{\rm d}t} = [f, H] + \frac{\partial f}{\partial t}.
\end{eqnarray}
If $H$ does not explicitly depend on time, the condition
\begin{eqnarray}
[f, H] = 0
\end{eqnarray}
implies that $f$ is an invariant of the motion
and therefore the Poisson brackets determine invariants of the motion.

\subsection{Lie operators}
\subsubsection{Definition}
We can define
\begin{eqnarray}
 :f:g = [f, g],
\label{eq:35}
\end{eqnarray}
where $:f:$ is an operator acting on the function $g$:
\begin{eqnarray}
 :f: = [f,  ].
\label{eq:36}
\end{eqnarray}
The operator {{$:f:$}} is called a {{Lie operator}}.

We could say colloquially that Lie operators are Poisson brackets in waiting. A Lie operator  acts on functions {{$g(x, p)$}}, and we have special cases where the function $g$ itself is a canonical variable:
\begin{eqnarray}
 g(x, p) = x \rightarrow :f:x,
\label{eq:37}
\end{eqnarray}
\begin{eqnarray}
 g(x, p) = p \rightarrow :f:p.
\label{eq:38}
\end{eqnarray}

\subsubsection{Useful formulas for calculations}
With $x$ the coordinate and $p$ the momentum, we can write useful and frequently used expressions for special cases of $f$. From (\ref{eq:34}) and (\ref{eq:36}), it follows  that
\begin{eqnarray}
:x: = \frac{\partial}{\partial p}, \quad :p: = -\frac{\partial}{\partial x},
\end{eqnarray}
\begin{eqnarray}
:x:^{2} = \frac{\partial^{2}}{\partial p^{2}}, \quad :p:^{2} = \frac{\partial^{2}}{\partial x^{2}},
\end{eqnarray}
\begin{eqnarray}
:x^{2}: = 2x\frac{\partial}{\partial p}, \quad :p^{2}: = -2p\frac{\partial}{\partial x},
\end{eqnarray}
\begin{eqnarray}
:xp: = p\frac{\partial}{\partial p} - x\frac{\partial}{\partial x}, \quad
:x::p: = :p::x: = -\frac{\partial^{2}}{\partial x \partial p}.
\end{eqnarray}
Applying those to the coordinate $x$ and the momentum $p$, we find for later use
\begin{eqnarray}
:p^{2}:x = \frac{\partial p^{2}}{\partial x}\frac{\partial x}{\partial p} - \frac{\partial p^{2}}{\partial p}\frac{\partial x}{\partial x} = -2p,
\end{eqnarray}
\begin{eqnarray}
:p^{2}:p = \frac{\partial p^{2}}{\partial x}\frac{\partial p}{\partial p} - \frac{\partial p^{2}}{\partial p}\frac{\partial p}{\partial x} = 0,
\end{eqnarray}
\begin{eqnarray}
(:p^{2}:)^{2}x = :p^{2}:(:p^{2}:x) = :p^{2}:(-2p) = 0,
\end{eqnarray}
\begin{eqnarray}
(:p^{2}:)^{2}p = :p^{2}:(:p^{2}:p) = :p^{2}:(0) = 0.
\end{eqnarray}

\subsection{Lie transformations}
We can define powers of Lie operators as
\begin{eqnarray}
 (:f:)^{2}g = :f:(:f:g) = [f,[f, g]], \ {\mathrm{etc.}}
\end{eqnarray}
and in particular extend this to a series of the form
\begin{eqnarray}
 e^{:f:}  = \sum_{i=0}^{\infty} \frac{1}{i!}(:f:)^{i},
\label{eq:39}
\end{eqnarray}
\begin{eqnarray}
 e^{:f:}  = 1 + :f: + \frac{1}{2!}(:f:)^{2} + \frac{1}{3!}(:f:)^{3} + \cdots.
\label{eq:40}
\end{eqnarray}
The operator {{$e^{:f:}$}} is called a {{Lie transformation}} and $f$ the generator of the
transformation. The generator $f$ can be any function of $x$ and $p$.


\subsubsection{Examples}
The following examples are mostly in one dimension for simplicity and easy reading, but the concept applies to full three-dimensional problems. We use as a first example the generator {{$f = -Lp^{2}/2$}}~and, using (\ref{eq:39}), we get
\begin{eqnarray}
{{e^{:-Lp^{2}/2:}}}x &= &x - \frac{1}{2}L{\underbrace{:p^{2}:x}_{= -2p}} + \frac{1}{8}L^{2}{\underbrace{(:p^{2}:)^{2}x}_{= 0}} + \cdots\\
                 &= &{x + Lp},\\
{{e^{:-Lp^{2}/2:}}}p &= &p - \frac{1}{2}L{\underbrace{:p^{2}:p}_{= 0}} + \cdots\\
                 &= &{p}.
\label{eq:46}
\end{eqnarray}
We have used some of the expressions derived before.

We find that Eq.~(\ref{eq:46}) is the transformation of a drift space of length $L$.

In general, we can say that a Lie transformation acting on the phase-space coordinates $x$ and $p$ at a location 1 propagates them to another location 2:
\begin{eqnarray}
{{e^{:f:}}} (x, p)_{1} = (x, p)_{2},
\end{eqnarray}
that is,
\begin{eqnarray}
{{e^{:f:}}} x_{1} = x_{2},
\label{eq:47a}
\end{eqnarray}
\begin{eqnarray}
{{e^{:f:}}} p_{1} = p_{2}.
\label{eq:47b}
\end{eqnarray}
The meaning of (\ref{eq:47a}) and (\ref{eq:47b}) is as follows.
\begin{itemize}
\item[i)]Lie transformations describe how to go from one point to another~\cite{bib:dragt, bib:ad01, bib:ad02, bib:ad03}.
\item[ii)]In our definition, it is a map from location 1 to location 2.
\item[iii)]The motion through a machine element (drift, magnet, etc.) is fully described by $f$.
\end{itemize}

\subsubsection{Generators of Lie transformations}
We can ask the question: what is the physical meaning of $f$?
\begin{itemize}
\item[i)]The generator $f$ is related to the Hamiltonian of the element.
\item[ii)]We use the Hamiltonian to describe the motion through an individual element.
\item[iii)]Inside a single element the motion is `smooth', while in the full machine it is not.
\item[iv)]Given an element of length $L$ and the Hamiltonian $H$, we get the generator $f = L\cdot H$.
\item[v)]We can track `thick' elements and still remain symplectic.
\end{itemize}
This forms our `local' description of the objects in an accelerator, independent of other machine elements or the global behaviour.

\subsubsection{Some formulas for Lie transformations}
Here we provide some useful formulas using Lie transformations which are frequently used in calculations. With $a$ being a constant, and $f, g$ arbitrary functions of the canonical variables $x$ and $p$,
\begin{eqnarray}
:a: = 0, \quad e^{:a:} = 1,
\end{eqnarray}
\begin{eqnarray}
:f:a = 0, \quad e^{:f:}a = a, \quad e^{:af:} = e^{a:f:},
\end{eqnarray}
\begin{eqnarray}
e^{:f:}g(x) = g(e^{:f:}x),
\end{eqnarray}
\begin{eqnarray}
e^{:f:}G(:g:)e^{-:f:} = G(:e^{:f:}g:),
\end{eqnarray}
\begin{eqnarray}
e^{:f:}[g, h] = [e^{:f:}g,e^{:f:}h],
\end{eqnarray}
\begin{eqnarray}
(e^{:f:})^{-1} = e^{-:f:}
\end{eqnarray}
and, very importantly,
\begin{eqnarray}
e^{:f:}e^{:g:}e^{-:f:} = e^{:e^{:f:}g:}.
\end{eqnarray}

\subsubsection{Some examples for Lie transformations}
For the generator
\begin{eqnarray}
f = -\frac{L}{2}p^{2},
\end{eqnarray}
using the formulas above, we obtain easily
\begin{eqnarray}
{{e^{:f:}}}x &= &x + Lp,\\
{{e^{:f:}}}p &= &p.
\end{eqnarray}
This is the map for a drift space.

For another important generator of the form
\begin{eqnarray}
f = -\frac{L}{2} (kx^{2} + p^{2}),
\end{eqnarray}
we have the map
\begin{eqnarray}
e^{:f:}x &= &e^{:-\frac{L}{2} (kx^{2} + p^{2}):}x,  \\
e^{:f:}p &= &e^{:-\frac{L}{2} (kx^{2} + p^{2}):}p.
\end{eqnarray}
For its evaluation, we should remember
\begin{eqnarray}
 e^{:f:}g = \sum_{n=0}^{\infty} \frac{:f:^{n}}{n!} g.
\end{eqnarray}

We get (without proof)
\begin{eqnarray}
e^{:-\frac{L}{2} (kx^{2} + p^{2}):}x &= &\sum_{n=0}^{\infty} \left( \frac{(-kL^{2})^{2n}}{(2n)!}\cdot x + L\frac{(-k
L^{2})^{2n+1}}{(2n+1)!}\cdot p\right), \\
e^{:-\frac{L}{2} (kx^{2} + p^{2}):}p &= &\sum_{n=0}^{\infty} \left( \frac{(-kL^{2})^{2n}}{(2n)!}\cdot p - k\frac{(-kL^{2})^{2n+1}}{(2n+1)!}\cdot x\right).
\end{eqnarray}
The terms within the sums correspond to the trigonometric functions sine and cosine.
Therefore, the generator
\begin{eqnarray}
f = -\frac{L}{2} (kx^{2} + p^{2})
\end{eqnarray}
describes the map
\begin{eqnarray}
e^{:f:}x &= &x \cos(\sqrt{k}L) + \frac{p}{\sqrt{k}} \sin(\sqrt{k}L),\\
e^{:f:}p &= &-\sqrt{k} x~{\mathrm{sin}}(\sqrt{k}L) + p \cos (\sqrt{k}L),
\end{eqnarray}
which is the map for a thick, focusing quadrupole in one dimension.

\subsubsection{Hamiltonians for accelerator elements}
We can extend the concept to a full Hamiltonian in three dimensions and
the generators for the most common magnets in three dimensions are:
dipole,
\begin{eqnarray}
H = -\frac{-x \delta}{\rho} + \frac{x^{2}}{2\rho^{2}} + \frac{p_{x}^{2} + p_{y}^{2}}{2(1 + \delta)};
\label{eq:122}
\end{eqnarray}
quadrupole,
\begin{eqnarray}
H = \frac{1}{2}k_{1}(x^{2} - y^{2}) + \frac{p_{x}^{2} + p_{y}^{2}}{2(1 + \delta)};
\label{eq:123}
\end{eqnarray}
sextupole,
\begin{eqnarray}
H = \frac{1}{6}k_{2}(x^{3} - 3 x y^{2}) + \frac{p_{x}^{2} + p_{y}^{2}}{2(1 + \delta)};
\label{eq:124}
\end{eqnarray}
octupole,
\begin{eqnarray}
H = \frac{1}{4}k_{3}(x^{4} - 6 x^{2} y^{2} + y^{4}) + \frac{p_{x}^{2} + p_{y}^{2}}{2(1 + \delta)}.
\label{eq:125}
\end{eqnarray}

In many cases the non-linear effects by the kinematic term are negligible $(\delta \ll 1)$ and
the Hamiltonian for a quadrupole
\begin{eqnarray}
H = \frac{1}{2}k_{1}(x^{2} - y^{2}) + \frac{p_{x}^{2} + p_{y}^{2}}{2(1 + \delta)}
\end{eqnarray}
can be simplified to
\begin{eqnarray}
H = \frac{1}{2}k_{1}(x^{2} - y^{2}) + \frac{p_{x}^{2} + p_{y}^{2}}{2}.
\end{eqnarray}
In one dimensions (without $y$ and $p_{y}$), this reduces to the previous example.

\subsubsection{The use of this concept}
If we know the Hamiltonian of a machine element (e.g.\ a magnet), then
\begin{eqnarray}
e^{:LH:} x_{0} = x_{1},
\end{eqnarray}
\begin{eqnarray}
e^{:LH:} p_{0} = p_{1}.
\end{eqnarray}
This is also true for any function of $x$ and $p$:
\begin{eqnarray}
e^{:LH:} f_{0}(x, p) = f_{1}(x, p).
\end{eqnarray}
Since we are now interested in the transformation of the coordinates and momenta from one location
to the next, we have to multiply the Hamiltonian (invariant) with the length $L$ of the element.

The miracles of this method are as follows.
\begin{itemize}
\item[i)] Poisson brackets always create {\textit{symplectic}} maps.
\item[ii)] The exponential form $e^{:h:}$ is {\textit{always}} symplectic.
\item[iii)] Better: the exponent is directly connected to the invariant of the transfer map.
\end{itemize}

\subsubsection{OTM and effective Hamiltonian}
We can combine many machine elements $f_{n}$ by applying one
transformation after the other:
\begin{eqnarray}
e^{:h:} = e^{:f_{1}:} e^{:f_{2}:} \ldots
e^{:f_{N}:}.
\label{eq:30}
\end{eqnarray}
We note the difference to the earlier treatment with linear elements:
\begin{itemize}
\item[i)] not restricted to matrices, i.e.\ linear elements;
\item[ii)] we arrive at a transformation for the full ring $\rightarrow$ a OTM;
\item[iii)] the OTM is the exponential of the effective Hamiltonian.
\end{itemize}
It can be written as
\begin{eqnarray}
{\cal{M}}_{\rm ring} =e^{:{{-C}} {\cal{H}}_{\rm eff}:},
\end{eqnarray}
where $C$ is the circumference of the whole ring and the effective Hamiltonian is the
invariant.

\subsection{Campbell--Baker--Hausdorff formula}
We have to find an algorithm to combine elements in the form (\ref{eq:30}).

In the case where the two functions $f$ and $g$ commute (i.e.\ $[f,g] = [g,f]$),
the concatenation is very easy:
\begin{eqnarray}
e^{:h:} = e^{:f:} e^{:g:} = e^{:f + g:}.
\end{eqnarray}
In all other cases
we can use the Campbell--Baker--Hausdorff (CBH) formula~\cite{bib:haus,bib:camp,bib:poin02,bib:baker}) to obtain
the effective Hamiltonian $h$ (or OTM) and therefore
\begin{eqnarray}
\begin{array}{ll}
h = f & +  g  +  \frac{1}{2}[f,g]  +  \frac{1}{12}[f,[f,g]]  +  \frac{1}{12}[g,[g,f]] \\
    & +  \frac{1}{24}[f,[g,[g,f]]]  -  \frac{1}{720}[g,[g,[g,[g,f]]]]\\
    & -  \frac{1}{720}[f,[f,[f,[f,g]]]]  +  \frac{1}{360}[g,[f,[f,[f,g]]]]  +  \cdots \\
{\mathrm{or}}~&~ \\
h = f & +  g  +  \frac{1}{2}:f:g  +  \frac{1}{12}:f:^{2}g  +  \frac{1}{12}:g:^{2}f \\
    & +  \frac{1}{24}:f::g:^{2}f  -  \frac{1}{720}:g:^{4}f \\
    & - \frac{1}{720}:f:^{4}g  +  \frac{1}{360}:g::f:^{3}g  +  \cdots.
\end{array}
\end{eqnarray}
If we proceed for all machine elements, we obtain the non-linear, effective Hamiltonian, corresponding to the OTM.

We have a special case when we combine
\begin{eqnarray}
e^{:h:} = e^{:f:} e^{:g:}
\end{eqnarray}
if one of them ($f$ or $g$) is small.
Then we can truncate the series and get a very useful formula. Assume that $g$ is small:
\begin{eqnarray}
e^{:f:} e^{:g:} = e^{:h:} = \exp \left[:f + \left( \frac{:f:}{1 - e^{-:f:}}\right) g
+ {\cal{O}}(g^{2}): \right].
\end{eqnarray}

\subsection{General non-linear elements}
For a general thin-lens kick given by $f(x)$, we write for the Lie transformation
\begin{eqnarray}
e^{:\int _{0}^{x} f(u) \, {\mathrm{d}}u:},
\end{eqnarray}
which gives for the map in one dimension
\begin{eqnarray}
\begin{array}{lll}
 x &= &x_{0}, \\
 p &= &p_{0} + f(x).\\
\end{array}
\end{eqnarray}
For example, a thin-lens multipole of order $n$ (i.e. $f(x) = a\cdot x^{n}$):
\begin{eqnarray}
e^{:\frac{a}{n+1}\cdot x^{n+1}:}
\end{eqnarray}
gives for the map
\begin{eqnarray}
\begin{array}{lll}
 x &= &x_{0}, \\
 p &= &p_{0} + a x^{n}. \\
\end{array}
\end{eqnarray}

Monomials in $x$ and $p$ of orders $n$ and $m$ $(x^{n}p^{m})$
\begin{eqnarray}
e^{:a x^{n}p^{m}:}
\end{eqnarray}
give for the map (for $n \neq m$)
\begin{eqnarray}
\begin{array}{lll}
 e^{:a x^{n}p^{m}:} x  &= &x \cdot [ 1 + a(n - m)x^{n-1}p^{m-1}]^{m/(m-n)},\\
 e^{:a x^{n}p^{m}:} p  &= &p \cdot [ 1 + a(n - m)x^{n-1}p^{m-1}]^{n/(n-m)},\\
\end{array}
\end{eqnarray}
and for the special map (for $n = m$)
\begin{eqnarray}
\begin{array}{lll}
 e^{:a x^{n}p^{n}:} x  &= &x \cdot e^{-a nx^{n-1}p^{n-1}},\\
 e^{:a x^{n}p^{n}:} p  &= &p \cdot e^{a nx^{n-1}p^{n-1}}.\\
\end{array}
\end{eqnarray}

\subsection{From Hamiltonian to maps}
We have seen that given the Hamiltonian ${{H}}$ of a machine element
of length $L$, the Lie operator becomes
\begin{eqnarray}
 f \rightarrow :f: = :L H: .
\end{eqnarray}
The corresponding map for an element with length $L$ is then
\begin{eqnarray}
  e^{:f:} = e^{:{{-L}} H:}.
\end{eqnarray}
Transformations derived from a Hamiltonian are always symplectic
and we have (in one dimension)
\begin{eqnarray}
{{e^{:f:}}} x_{0} = x_{1},
\end{eqnarray}
\begin{eqnarray}
{{e^{:f:}}} p_{0} = p_{1}
\end{eqnarray}
or, using $Z = (x, p_{x}, y, p_{y}, \ldots)$ (in two dimensions),
\begin{eqnarray}
{{e^{:f:}}} Z_{0} = Z_{1}.
\end{eqnarray}

\subsection{From maps to Hamiltonian}
Assuming we do not have the Hamiltonian,
but a matrix $\mathbf{M}$ (from somewhere, including a numerical map from a tracking code),
we search for the Hamiltonian.
Starting from the well-known one-turn matrix
\begin{eqnarray}
{{\cal{M}}} \equiv
\left( \begin{array}{cc}
\cos(\mu) + \alpha \sin(\mu)  &\beta \sin(\mu) \\
-{\gamma}\sin(\mu) &\cos(\mu) - \alpha \sin(\mu)\\
\end{array}\right)
\end{eqnarray}
or, rewritten,
\begin{eqnarray}
{{\cal{M}}} Z_{0} = Z_{1},
\end{eqnarray}
we search for the corresponding form of ${{f}}$, such as
\begin{eqnarray}
{{\cal{M}}} \leftrightarrow e^{:f:} \quad ( = e^{:{{-\mu}} H:}).
\end{eqnarray}

For the linear matrix, we know that $f$ must be a {{quadratic}} form in $x$ and $p$.\\
Any quadratic form can be written as
\begin{eqnarray}
 f = -\frac{1}{2}Z^{*} F Z \quad \left[ = -\frac{1}{2} (a\cdot x^{2} + b\cdot xp + c\cdot p^{2})\right],
\end{eqnarray}
where $F$ is a symmetric, positive-definite matrix.

Then we can write (without proof, see e.g. \cite{bib:dragt})
\begin{eqnarray}
 :f: Z = S F Z,
\end{eqnarray}
where $S$ is the `symplecticity' matrix.

Therefore, we get for the Lie transformation
\begin{eqnarray}
 e^{:f:} Z \leftrightarrow e^{SF} Z.
\end{eqnarray}

Since we have $n = 2$, we get (using the Hamilton--Cayley theorem)
\begin{eqnarray}
 e^{SF} = \exp \left( \begin{array}{cc}
b  &c \\
-a  &-b \\
\end{array}\right)
= a_{0} + a_{1} \left( \begin{array}{cc}
b  &c \\
-a  &-b \\
\end{array}\right).
\end{eqnarray}
We now have to find $a_{0}$ and $a_{1}$.
The eigenvalues of $SF$ are
\begin{eqnarray}
  \lambda_{\pm} = \pm {\rm i} \sqrt{ac - b^{2}}.
\end{eqnarray}

This tells us for the coefficients the conditions
\begin{eqnarray}
 e^{\lambda_{+}} = a_{0} + a_{1}\cdot\lambda_{+},
\end{eqnarray}
\begin{eqnarray}
 e^{\lambda_{-}} = a_{0} + a_{1}\cdot\lambda_{-}
\end{eqnarray}
and, therefore,
\begin{eqnarray}
a_{0} = \cos (\sqrt{ac - b^{2}}),
\end{eqnarray}
\begin{eqnarray}
a_{1} = \frac{ \sin (\sqrt{ac - b^{2}})}{\sqrt{ac - b^{2}}}
\end{eqnarray}
and
\begin{eqnarray}
e^{SF} = \cos (\sqrt{ac - b^{2}}) + \frac{\sin (\sqrt{ac - b^{2}})}{\sqrt{ac - b^{2}}}
\left( \begin{array}{cc}
b  &c \\
-a  &-b \\
\end{array}\right).
\end{eqnarray}

For a general $2 \times 2$ matrix,
\begin{eqnarray}
 M = \left( \begin{array}{cc}
m_{11}  &m_{12} \\
m_{21}  &m_{22} \\
\end{array}\right),
\end{eqnarray}
we get by comparison
\begin{eqnarray}
\cos (\sqrt{ac - b^{2}}) = \frac{1}{2} \mathrm{tr}(M)
\end{eqnarray}
and
\begin{eqnarray}
\frac{a}{-m_{21}} = \frac{2b}{m_{11} - m_{22}} = \frac{c}{m_{12}}
= \frac{\sqrt{ac - b^{2}}}{\sin (\sqrt{ac - b^{2}})}
\end{eqnarray}
for the quadratic form of ${{f}}$:
\begin{eqnarray}
f = -\frac{1}{2}(a\cdot x^{2} + b\cdot xp + c\cdot p^{2}).
\end{eqnarray}

\subsubsection{Examples}
For the example of a drift space with length $L$,
\begin{eqnarray}
{{\cal{M}}} \equiv
\left( \begin{array}{cc}
1  &L  \\
0 &1 \\
\end{array}\right),
\end{eqnarray}
we find
\begin{eqnarray}
 a = 0, \quad b = 0, \quad c = L
\end{eqnarray}
and, therefore, for the generator:
\begin{eqnarray}
f = -\frac{1}{2}(L p^{2}).
\end{eqnarray}

For the example of a thin quadrupole,
\begin{eqnarray}
{{\cal{M}}} \equiv
\left( \begin{array}{cc}
1  &0  \\
-\frac{1}{F} &1 \\
\end{array}\right),
\end{eqnarray}
we find
\begin{eqnarray}
 a = \frac{1}{F}, \quad b = 0, \quad c = 0
\end{eqnarray}
and for the generator:
\begin{eqnarray}
f = -\frac{1}{2F}(x^{2}).
\end{eqnarray}
Here I have used $F$ for the focal length to avoid confusion with the
generator $f$.

A very important example in one dimension is again the one-turn matrix:
\begin{eqnarray}
{{\cal{M}}} \equiv
\left( \begin{array}{cc}
\cos{\mu} + \alpha \sin(\mu)  &\beta \sin{\mu} \\
-{\gamma}\sin{\mu} &\cos{\mu} - \alpha \sin(\mu)\\
\end{array}\right).
\end{eqnarray}
This corresponds to
\begin{eqnarray}
e^{:f_{2}:} = e^{:-\mu {{\frac{1}{2}(\gamma x^{2} + 2 \alpha x p + \beta p^{2})}}:}.
\end{eqnarray}
In this form, $f$ is $-\mu \cdot (\mbox{Courant--Snyder invariant})$
\begin{eqnarray}
e^{:f_{2}:} = e^{:-\mu {{\epsilon}}:}.
\end{eqnarray}
We write the generator as $f_{2}$ as a special case of a transformation.
We have obtained a standard Lie transformation ($e^{:f_{2}:}$)
for the linear one-turn matrix, i.e.\ a rotation.

After using the normalized variables for a pure rotation,
\begin{eqnarray}
\left( \begin{array}{cc}
\cos{\mu} + \alpha \sin(\mu)  &\beta \sin{\mu} \\
-{\gamma}\sin{\mu} &\cos{\mu} - \alpha \sin(\mu)\\
\end{array}\right)
\quad \Rightarrow \quad
\left( \begin{array}{cc}
\cos{\mu}   &\sin{\mu} \\
-\sin{\mu} &\cos{\mu} \\
\end{array}\right),
\end{eqnarray}
\begin{eqnarray}
e^{:-\mu {{\frac{1}{2}(\gamma x^{2} + 2 \alpha x p + \beta p^{2})}}:}
\quad \Rightarrow \quad
e^{:-\mu {{\frac{1}{2}(x^{2} + p^{2})}}:}
\end{eqnarray}
and extending this to a three-dimensional linear system, we have for {{$f_{2}$}}
\begin{eqnarray}
f_{2} = -\frac{\mu_{x}}{2}(x^{2} + p_{x}^{2}) - \frac{\mu_{y}}{2}(y^{2} + p_{y}^{2})
- \frac{1}{2}\alpha_{c}\delta^{2}
\end{eqnarray}
or, using the invariant action variables {{$J$}},
\begin{eqnarray}
f_{2} = - \mu_{x}J_{x} - \mu_{y}J_{y} - \frac{1}{2}\alpha_{c}\delta^{2}.
\end{eqnarray}
This is very useful in all calculations as a standard ($e^{:f_{2}:}$) transformation in three dimensions.

\section{Non-linear normal forms}
Normal-form transformations can be generalized for non-linear maps (i.e.\ not matrices).
If {{$\cal{M}$}} is our usual OTM, we try to find a transformation:
\begin{eqnarray}
{\cal{{{N}}  =  {{A}} {{M}} {{A}}}}^{-1},
\end{eqnarray}
where ${\cal{{{N}}}}$ is a simple form (like the rotation we had before).
Of course now we do not have matrices; we use a Lie transform with the
generator ${{F}}$ to describe the transformation {{$\cal{A}$}}:
\begin{eqnarray}
{\cal{{{N}}}}  = {{e^{-:h:}}} =  {\cal{{{A}} {{M}} {{A}}}}^{-1} =  {{e^{:F:}}} {\cal{{M}}
} {{e^{-:F:}}}.
\end{eqnarray}
The canonical transformation {{$\cal{A}$}}:
\begin{eqnarray}
{\cal{{{N}}  =  {{A}} {{M}} {{A}}}}^{-1} \quad \Rightarrow \quad {{{\cal{A}} = e^{:F:} }}
\end{eqnarray}
should be the transformation to produce our simple form. The simple form should contain only the effective Hamiltonian and depend only on the action variables because it is an invariant:
\begin{itemize}
\item[i)] $h(J_{x}, \Psi_{x}, J_{y},\Psi_{y}, z, \delta)~~\Rightarrow~~h(J_{x}, J_{y},\delta) = h_{\rm eff}(J_{x}, J_{y},\delta)$;
\item[ii)] Should work for any kind of local perturbation;
\item[iii)] Formalism and software tools exist to find $F$~(see e.g.~\cite{bib:chao, bib:ef05});
\item[iv)] Once we know $h_{\rm eff}(J_{x}, J_{y},\delta)$, we can derive everything.
\end{itemize}
Once we can write the map as (now in three dimensions)
\begin{eqnarray}
 {\cal{{{N}}}} = e^{-:h_{\rm eff}(J_{x}, J_{y},\delta):},
\end{eqnarray}
where $h_{\rm eff}$ depends only on $J_{x}, J_{x}$ and $\delta$, then we can immediately get the tunes:
\begin{eqnarray}
Q_{x}(J_{x}, J_{y},\delta) = \frac{1}{2\pi}\frac{\partial h_{\rm eff}}{\partial J_{x}},
\label{eq:42}
\end{eqnarray}
\begin{eqnarray}
Q_{y}(J_{x}, J_{y},\delta) = \frac{1}{2\pi}\frac{\partial h_{\rm eff}}{\partial J_{y}}
\label{eq:43}
\end{eqnarray}
and the change of path length:
\begin{eqnarray}
\Delta z = -\frac{\partial h_{\rm eff}}{\partial \delta}.
\end{eqnarray}
Particles with different $J_{x}, J_{y}$ and $\delta$ have different tunes. The dependences of $Q_{x}$ and $Q_{y}$ on $J$ are the amplitude detunings; the dependences on $\delta$ are the chromaticities.

\subsection{Effective Hamiltonians}
The effective Hamiltonian can be written in a general form (here to third order)
(see e.g. \cite{bib:ef04}) as
\begin{eqnarray}
 h_{\rm eff} & = & \mu_{x}J_{x} + \mu_{y}J_{y} + \frac{1}{2}\alpha_{c}\delta^{2} \notag \\
 &&+ c_{x1} J_{x}\delta + c_{y1} J_{y}\delta + c_{3}\delta^{3} \notag \\
 &&+ c_{xx} J_{x}^{2} + c_{xy} J_{x}J_{y} + c_{yy} J_{y}^{2} + c_{x2} J_{x} \delta^{2} + c_{y2} J_{y} \delta^{2} + c_{4} \delta^{4}
\label{eq:43a}
\end{eqnarray}
and then (see (\ref{eq:42}) and (\ref{eq:43}))
\begin{eqnarray}
Q_{x}(J_{x}, J_{y},\delta)  = \frac{1}{2\pi}\frac{\partial h_{\rm eff}}{\partial J_{x}}
= \frac{1}{2\pi}\left( {{\mu_{x}}} + {{2c_{xx} J_{x} + c_{xy} J_{y}}} + {{c_{x1} \delta + c_{x2} \delta^{2}}} \right),\\
Q_{y}(J_{x}, J_{y},\delta)  = \frac{1}{2\pi}\frac{\partial h_{\rm eff}}{\partial J_{y}}
= \frac{1}{2\pi}\left( {{\mu_{y}}} + {{2c_{yy} J_{y} + c_{xy} J_{x}}} + {{c_{y1} \delta + c_{y2} \delta^{2}}} \right),
\end{eqnarray}
\begin{eqnarray}
 \Delta z = -\frac{\partial h_{\rm eff}}{\partial \delta} = \alpha_{c}\delta + 3c_{3}\delta^{2} + 4c_{4}\delta^{3} + c_{x1} J_{x} + c_{y1} J_{y} + 2c_{x2} J_{x} \delta + 2c_{y2} J_{y} \delta.
\end{eqnarray}
The physical meanings of the different coefficients we have obtained are:
\begin{itemize}
\item[i)] $\mu_{x}, \mu_{y}$: horizontal and vertical tunes;
\item[ii)] $\frac{1}{2}\alpha_{c}, c_{3}, c_{4}$: linear and non-linear `momentum compactions';
\item[iii)] $c_{x1}, c_{y1}$: first-order chromaticities;
\item[iv)] $c_{x2}, c_{y2}$: second-order chromaticities;
\item[v)] $c_{xx}, c_{xy}, c_{yy}$: detuning with amplitude.
\end{itemize}

\subsubsection{Example: sextupole}
For the sextupole, we use the Hamiltonian (\ref{eq:124}) in one dimension. With the effective Hamiltonian we get the physical quantities of interest up to second order:
\begin{eqnarray}
 {\cal{{M}}} = e^{-:\frac{\mu}{2}x^{2} + p^{2} + \frac{1}{2}\alpha_{c}\delta^{2}:}
 e^{:f_{3}:} = e^{-:\mu J_{x} + \frac{1}{2}\alpha_{c}\delta^{2}:}
 e^{:\frac{k}{6} x^{3} + \frac{p^{2}}{2(1+\delta)}:}.
\end{eqnarray}
We get for $h_{\rm eff}$ (see e.g. \cite{bib:chao,bib:forest})
\begin{eqnarray}
 h_{\rm eff} =  \mu_{x} J_{x} + \frac{1}{2} \alpha_{c}\delta^{2} - \frac{k}{6} D^{3}\delta^{3} - 3 \frac{k}{6} \beta_{x} J_{x} D \delta
\end{eqnarray}
or, in three dimensions,
\begin{eqnarray}
 h_{\rm eff} =  \mu_{x} J_{x} + \mu_{y} J_{y} + \frac{1}{2} \alpha_{c}\delta^{2} - k D^{3}\delta^{3} - 3 \frac{k}{6} \beta_{x} J_{x} D \delta + 3 \frac{k}{6} \beta_{y} J_{y} D \delta.
\end{eqnarray}

When we have $h_{\rm eff}$ in three dimensions, we obtain
(see (\ref{eq:42}) and (\ref{eq:43}))
\begin{eqnarray}
Q_{x}(J_{x}, J_{y},\delta) = \frac{1}{2\pi}\frac{\partial h_{\rm eff}}{\partial J_{x}} = \frac{1}{2\pi} ({{\mu_{x}}} - {{3
\frac{k}{6} \beta_{x} D \delta}}),
\end{eqnarray}
\begin{eqnarray}
Q_{y}(J_{x}, J_{y},\delta) = \frac{1}{2\pi}\frac{\partial h_{\rm eff}}{\partial J_{y}} = \frac{1}{2\pi} ({{\mu_{y}}} + {{3
\frac{k}{6} \beta_{y} D \delta}})
\end{eqnarray}
and the change of path length
\begin{eqnarray}
\Delta s = -\frac{\partial h_{\rm eff}}{\partial \delta}
= \alpha_{c} \delta - 3 \frac{k}{6} D^{3}\delta^{2} - 3 \frac{k}{6} D(\beta_{x}J_{x} - \beta_{y
}J_{y}).
\end{eqnarray}
With the terms $({1}/{2 \pi})({k}/{2}) \beta D \delta$,
we have recovered the usual chromaticities from sextupoles.

\subsubsection{Example: octupole}
Assume a linear rotation (as always using our `standard map') followed by an octupole; the Hamiltonian is 1D to keep it simple, but the formulas are valid for higher dimensions):
\begin{eqnarray}
 {\cal{H}} = \frac{\mu}{2}(x^{2} + p^{2}) + k_{3}\cdot\frac{x^{4}}{4} \quad (p = p_{x}),
\end{eqnarray}
with Hamilton's equations leading to
\begin{eqnarray}
 \dot{x} = \frac{\partial {\cal{H}}}{\partial p} = \mu p,
\end{eqnarray}
\begin{eqnarray}
 \dot{p} = -\frac{\partial {\cal{H}}}{\partial x} = -\mu x - k_{3}\cdot x^{3}.
\end{eqnarray}
The map, written in Lie representation, is
\begin{eqnarray}
 {\cal{{M}}} = e^{-\frac{\mu}{2}:x^{2} + p^{2}:} e^{:k_{3}\cdot\frac{x^{4}}{4}:} = R e^{:k_{3}\cdot\frac{x^{4}}{4}:}.
\end{eqnarray}
We transform it by applying
\begin{eqnarray}
{\cal N} & = & {{A}} {{M}} {{A}}^{-1}  =  e^{:F:} R e^{:k_{3}\cdot\frac{x^{4}}{4}:} e^{-:F:} = R R^{-1} e^{:F:}  Re^{:k_{3}\cdot\frac{x^{4}}{4}:} e^{-:F:},\\
& = & R e^{:R^{-1}F + k_{3}\cdot\frac{x^{4}}{4} - F: + O(\epsilon^{2})} = R e^{:{{(R^{-1} - 1)F + k_{3}\cdot\frac{x^{4}}{4}}}: + O(\epsilon^{2})}.
\end{eqnarray}
We have now to choose $F$ to simplify the expression
\begin{eqnarray}
 {{(R^{-1} - 1)F + k_{3}\cdot\frac{x^{4}}{4}   }}
\end{eqnarray}
and get (e.g. \cite{bib:forest})
\begin{eqnarray}
F = -\frac{1}{64}\{-5x^{4} + 3p^{4} + 6x^{2}p^{2} + x^{3}p\cdot(8\cot(\mu) + 4\cot(2\mu))
+ xp^{3}(8\cot(\mu) - 4\cot(2\mu))\}.
\end{eqnarray}
We go back to $x$ and $p$ coordinates and, with
\begin{eqnarray}
 J = (x^{2} + p^{2})/2,
\end{eqnarray}
we can write the map to second order:
\begin{eqnarray}
 M = e^{-:F:} e^{:-\mu J + {{\frac{3}{8} k_{3}\cdot J^{2}}}:} e^{:F:}.
\end{eqnarray}
The term ${{\frac{3}{8} k_{3}\cdot J^{2}}}$ produces the tune shift with
amplitude we know for an octupole ($<\beta^{2}>$) in real space.
Note that the normalized map (our most simple map)
\begin{eqnarray}
 R = e^{:-\mu J + {{\frac{3}{8} k_{3}\cdot J^{2}}}:}
\end{eqnarray}
is again a rotation in phase space, but the rotation angle now depends on the amplitude {$J$}, as expected.

We shall look at a concrete calculation.
When we have $h_{\rm eff}$ in one dimensions for a single octupole (see before),
\begin{eqnarray}
h_{\rm eff} = -\mu J + {{\frac{3}{8} k_{3}\cdot J^{2}}},
\end{eqnarray}
\begin{eqnarray}
Q_{x}(J_{x}, J_{y}) = \frac{1}{2\pi}\frac{\partial h_{\rm eff}}{\partial J_{x}}
= -\frac{1}{2\pi}\mu_{x} + {{\frac{3}{8 \cdot 2\pi} k_{3} J_{x}}}
\end{eqnarray}
and, with normalization in real space (i.e.\ $\beta \neq 1$),
\begin{eqnarray}
\Delta Q_{x}(J_{x}, J_{y}) = \frac{3}{8\cdot 2\pi} k_{3} \langle \beta^{2}\rangle J_{x}.
\end{eqnarray}
For example, for $\beta = 300$ m, $k_{3} = 0.01$:
\begin{eqnarray}
\Delta Q_{x}(J_{x}, J_{y}) = 53.7 \cdot J_{x}.
\end{eqnarray}

\subsection{Example: beam--beam effects}
\begin{figure}[t]
\centering\includegraphics[height= 5.10cm,width=  5.40cm]{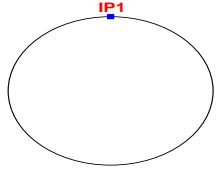}
\caption{Linear machine with a single interaction point (schematic)}
\label{fig:fig13}
\end{figure}

The beam--beam interaction in a particle collider is a very localized distortion, and a very strong non-linearity in the machine. The standard perturbation theory is not sufficient to treat this problem correctly. We show here how the Lie technique is applied in this case~\cite{bib:wh01}. Schematically, we show this in \Fref{fig:fig13}.

We are interested in the invariant {{$h$}}.
We need the (usual) linear transform $e^{:f_{2}:}$ and the
beam--beam interaction $e^{:F:}$, i.e.
\begin{eqnarray}
e^{:f_{2}:} \cdot e^{:F:} = e^{:{{h}}:}
\end{eqnarray}
with (see before)
\begin{eqnarray}
f_{2} =  -\frac{\mu}{2} \left( \frac{x^{2}}{\beta} + \beta p^{2}_{x}\right)
\end{eqnarray}
and (see before)
\begin{eqnarray}
F = \displaystyle{\int_{0}^{x}} \, {\mathrm{d}}x' f(x').
\end{eqnarray}

For a Gaussian beam, we have for $f(x)$ (see article on beam--beam effects~\cite{bib:tp01})
\begin{eqnarray}
f(x) = \frac{2}{x} ( 1 - e^{{-x^{2}}/{2\sigma^{2}}}).
\end{eqnarray}
As usual, we go to action-angle variables $\Psi$, $J$:
\begin{eqnarray}
x = \sqrt{2J\beta} \sin \Psi, \quad p = \sqrt{\frac{2J}{\beta}}\cos\Psi
\end{eqnarray}
and write $F(x)$ as a Fourier series:
\begin{eqnarray}
F(x) = \sum_{n=-\infty}^{\infty}  c_{n}(J) {\rm e}^{{\rm i}n\Psi}.
\end{eqnarray}

With this transformation to action-angle variables, the generator $f_{2}$ becomes very simple:
\begin{eqnarray}
f_{2} = -\mu J
\end{eqnarray}
and useful properties of Lie operators (e.g. \cite{bib:chao, bib:forest, bib:herr})
\begin{eqnarray}
:f_{2}:g(J) = 0, \quad :f_{2}:{\rm e}^{{\rm i}n\Psi} = {\rm i}n \mu {\rm e}^{{\rm i}n\Psi},
\quad g(:f_{2}:){\rm e}^{{\rm i}n \Psi} = g({\rm i}n \mu) {\rm e}^{{\rm i}n\Psi}
\end{eqnarray}
and the formula (e.g. \cite{bib:chao, bib:forest, bib:herr})
\begin{eqnarray}
e^{:f_{2}:} e^{:F:} = e^{:h:} = \exp \left[:f_{2} + \left( \frac{:f_{2}:}{1 - e^{-:f_{2}:}}\right)
F + {\cal{O}}(F^{2}): \right]
\end{eqnarray}
give immediately for $h$
\begin{eqnarray}
h = -\mu J + \sum_{n}  c_{n}(J) \frac{{\rm i} n \mu}{1 - {\rm e}^{-{\rm }{\rm i} n \mu}}
{\rm e}^{{\rm i}n\Psi},
\end{eqnarray}
\begin{eqnarray}
h = -\mu J + \sum_{n}  c_{n}(J) \frac{n \mu}{2 \sin ({n \mu}/{2})} {\rm e}^{\left({\rm i}n\Psi + {\rm i}\frac{n \mu}{2}\right)}.
\end{eqnarray}
Away from resonances, the normal-form transformation gives
\begin{eqnarray}
h_{n} = -\mu J + c_{0}(J) =\mathrm{const}.
\end{eqnarray}
The expression
\begin{eqnarray}
\frac{{\rm d} c_{0}(J)}{{\rm d} J}
\end{eqnarray}
gives the detuning with amplitude due to the beam--beam interaction.

The analysis for a single interaction point gives~\cite{bib:wh01}
\begin{eqnarray}
h = -\mu J + \sum_{n}  c_{n}(J) \frac{n \mu}{{{2 \sin ({n \mu}/{2})}}}
{\rm e}^{\left({\rm i}n\Psi + {\rm i}\frac{n \mu}{2}\right)}.
\end{eqnarray}
On resonance, we immediately find the conditions
\begin{eqnarray}
Q = \frac{p}{n} = \frac{\mu}{2 \pi}
\end{eqnarray}
with $c_{n} \neq 0$:
\begin{eqnarray}
\sin \left(\frac{n \pi p}{n}\right)
= \sin (p \pi) \equiv 0 \quad \mbox{for all integer} ~ p
\end{eqnarray}
and $h$ diverges, finding automatically all resonance conditions.

\begin{figure}[t]
\centering{{\includegraphics[height= 3.50cm,width=  4.50cm]{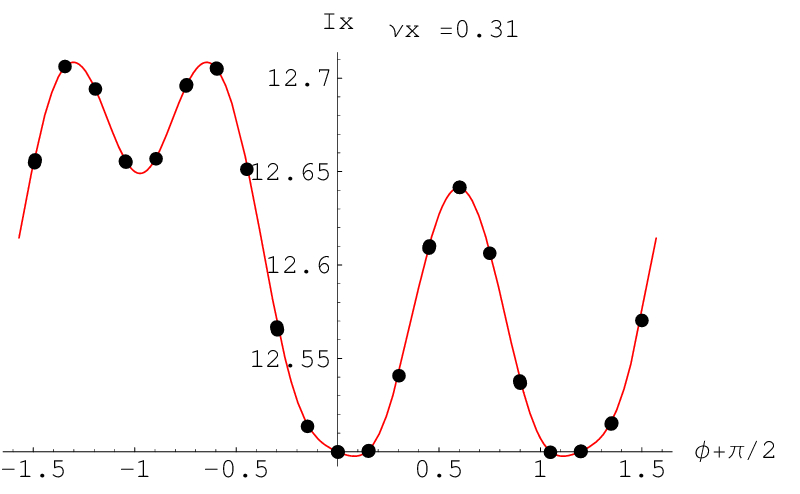}}
{\includegraphics[height= 3.50cm,width=  4.50cm]{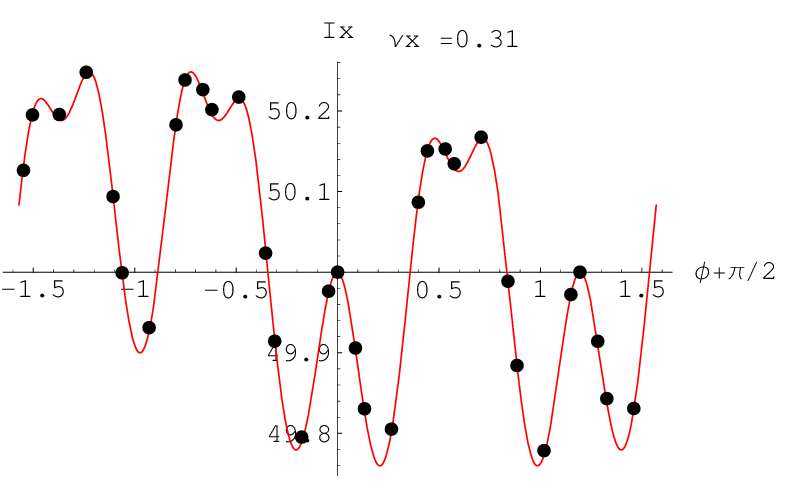}}}
\caption{Turn by turn points in phase space from simulation program, comparison with invariant~(full line).}
\label{fig:fig15}
\end{figure}

To compare with particle tracking, a simple tracking code was used~\cite{bib:wh01} and the turn by turn points in phase space are shown in \Fref{fig:fig15} for amplitudes of 5$\sigma_{x}$ and 10$\sigma_{x}$. The figure also shows the beam--beam invariant as computed from the effective Hamiltonian compared with the tracking result~\cite{bib:wh01}. The agreement is remarkable.

This technique can be used to analyse and compare the strengths of non-linear effects as demonstrated in~\cite{bib:dk01} and used to derive scaling laws for long range beam--beam effects in the LHC \cite{bib:assmann}.

\section{Differential algebra}
\subsection{Motivation}
Analytical formulas are usually difficult to derive and often rely on
simplifications and assumptions.
Although they may be fast to evaluate, they can only be used as guidance to
more precise calculations.
It is generally accepted that elaborate tracking codes provide the most reliable
description of a non-linear system.
Ideally, one should like to derive global quantities such as closed orbit, tunes
or detuning from the tracking codes.
The tracking of a complicated system relates the output {\it{numerically}} to the input
producing a numerical map and one might ask whether it is possible to relate
this output to global quantities in an easy way.
More precisely, can a code be used to provide input to a description or formula
such that these relate more directly to global quantities?
This would free us from approximate analytical calculations.
They would provide `analytical' formulas although they are not analytical
in the classical sense, which should be considered as symbolic~\cite{bib:forest}.
A typical example of such an analytical derivation would be a Taylor series derived
from a tracking code.

\subsection{Taylor series as analytical formulas}
\begin{figure}[t]
\centering{{\includegraphics[height= 3.00cm,width= 8.850cm]{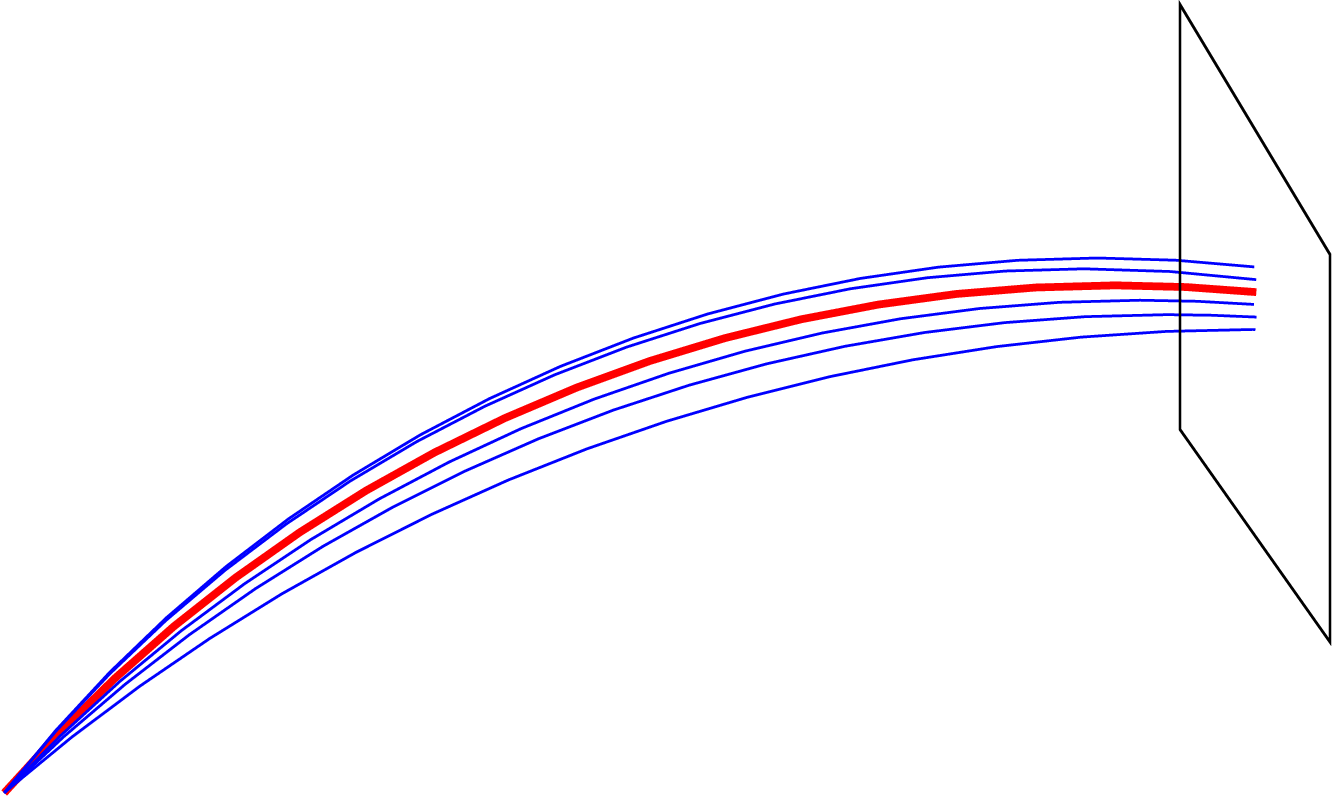}}}
\caption{Trajectories with small deviations from an ideal trajectory}
\label{fig:fig16}
\end{figure}

Why are Taylor series useful in the analysis of beam dynamics, in particular
stability for non-linear systems?
Let us study the paraxial behaviour sketched in \Fref{fig:fig16}.
The red line indicates the ideal orbit while the blue lines are small deviations from this
ideal orbit caused by distortions.
If we understand how small deviations behave, we understand the system much better.
Now remember the definition of the Taylor series:
\begin{eqnarray}
f(a + \Delta x) = f(a) + \sum_{n=1}^{\infty} \frac{f^{(n)}(a)}{n!} \Delta x^{n},
\end{eqnarray}
\begin{eqnarray}
f(a + \Delta x) = f(a) + \frac{{f'(a)}}{{{1!}}}\Delta x^{1} + \frac{{f''(a)}}{{{2!}}}\Delta x^{2} + \frac{{f'''(a)}}{{{3!}}}\Delta x^{3}  +\cdots.
\end{eqnarray}
The coefficients determine the behaviour of small deviations $\Delta x$ from the ideal orbit $x$, i.e.\ the Taylor expansion does a paraxial analysis of the system. If the function {{$f(x)$}} is represented by a Taylor series:
\begin{eqnarray}
f(a + \Delta x) = f(a) + \sum_{n=1}^{\infty} \frac{f^{(n)}(a)}{n!} \Delta x^{n}
\end{eqnarray}
and if it is truncated to the $m$th order:
\begin{eqnarray}
f(a + \Delta x) = f(a) + \sum_{n=1}^{m} \frac{f^{(n)}(a)}{n!} \Delta x^{n}.
\end{eqnarray}
What is the relevance?
\begin{itemize}
\item[i)]There is an equivalence between the function {{$f(x)$}} and the vector $(f(a), f'(a), f''(a), \ldots, f^{(m)}(a))$.
\item[ii)]This vector is a truncated power series algebra (TPSA) representation of $f(x)$ around {{$a$}}.
\end{itemize}
It can be shown that the global quantities of interest are obtained more easily and conveniently from the Taylor series map~\cite{bib:berz03}. Since it is derived from a symplectic tracking code, the corresponding map is also symplectic. Our problem is how to get these coefficients without extra work.

The problem is getting the derivatives $f^{(n)}(a)$ of $f(x)$ at $a$:
\begin{eqnarray}
f'(a) = \frac{f(a + \epsilon) - f(a)}{\epsilon}.
\end{eqnarray}
The problems with this approach are as follows.
\begin{itemize}
\item[i)] Need to subtract almost equal numbers and divide by small number.
\item[ii)] For higher orders $f'', f''', \ldots$, it is hopeless to obtain the required accuracy.
\item[iii)] We can use differential algebra (DA)~\cite{bib:berz01, bib:berz02, bib:berz03}.
\end{itemize}

\subsection{Basic concepts}
The basic idea of differential algebra is the automatic differentiation of results
produced by a tracking code to provide the coefficients of a Taylor series.
Here we give the basic concepts of the mathematics~\cite{bib:berz01, bib:berz02, bib:berz03}:
\begin{enumerate}
\item[1.] Define a pair ($q_{0}, q_{1}$),~~with $q_{0}, q_{1}$ real numbers.\\

\item[2.] Define operations on a pair like
\begin{eqnarray}
(q_{0}, q_{1}) + (r_{0}, r_{1}) = (q_{0} + r_{0}, q_{1} + r_{1}),\end{eqnarray}
\begin{eqnarray}
c \cdot (q_{0}, q_{1}) = (c\cdot q_{0}, c \cdot q_{1}),\end{eqnarray}
\begin{eqnarray}
(q_{0}, q_{1}) \cdot (r_{0}, r_{1}) = ( q_{0}\cdot r_{0}, q_{0}\cdot r_{1}
+ q_{1}\cdot r_{0}).
\end{eqnarray}

\item[3.] And some ordering:
\begin{eqnarray}
(q_{0},q_{1}) < (r_{0},r_{1}) \quad {\mathrm{if}}\ q_{0} < r_{0} \ {\mathrm{or}} \
(q_{0} = r_{0} \ {\mathrm{and}} \ q_{1} < r_{1}),\end{eqnarray}
\begin{eqnarray}
(q_{0},q_{1}) > (r_{0},r_{1}) \quad {\mathrm{if}} \ q_{0} > r_{0}
\ {\mathrm{or}} \ (q_{0} = r_{0} \ {\mathrm{and}} \ q_{1} > r_{1}).\end{eqnarray}

\item[4.] This implies something strange:
\begin{eqnarray}
(0,0) < (0,1) < (r,0) \ ({\mathrm{for~any~positive}}~r),
\end{eqnarray}
\begin{eqnarray}
(0, 1) \cdot (0, 1) = (0, 0) \rightarrow (0, 1) = \sqrt{(0, 0)}.
\end{eqnarray}

This means that (0, 1) is between 0 and \textit{any} real number
$\rightarrow$ infinitely small.
Therefore we call this a `differential unit' $d = (0,1) = \delta$.
\end{enumerate}

Of course $(q, 0)$ is just the real number $q$ and we define
the `real part' and `differential part'
(a bit like complex numbers):
\begin{eqnarray}
q_{0} = {\cal{R}}(q_{0},q_{1}) \quad {\mathrm{and}} \quad q_{1} = {\cal{D}}(q_{0},q_{1}).
\end{eqnarray}

With our rules, we can further see that
\begin{eqnarray} (1,0) \cdot (q_{0},q_{1}) = (q_{0},q_{1}), \end{eqnarray}
\begin{eqnarray} (q_{0},q_{1})~^{-1} = \left(\frac{1}{q_{0}},
-\frac{q_{1}}{q_{0}^{2}} \right). \end{eqnarray}

Now we let a function {{$f$}} act on the pair (or vector)
using our rules.
For example,
\begin{eqnarray}
  f(x) \rightarrow f(x,0)
\end{eqnarray}
acts like the function {{$f$}} on the real variable $x$:
\begin{eqnarray}
 f(x) = {\cal{R}}[f(x,0)].
\end{eqnarray}
What about the differential part {{$\cal{D}$}}?

For a function $f(x)$, without proof,
\begin{eqnarray}
{\cal{D}}[f(x + d)] = {\cal{D}}[f((x, 0) + (0, 1))] = {\cal{D}}[f(x, 1)] = f'(x).
\end{eqnarray}
The differential part of the output is the first-order derivative of the function $f(x)$.
To demonstrate this, it is done most easily by looking at an example.

Assume for the function $f(x)$:
\begin{eqnarray} f(x) = x^{2} + \frac{1}{x}; \end{eqnarray}
then, using school calculus, we can get the derivative:
\begin{eqnarray} f'(x) = 2x - \frac{1}{x^{2}}. \end{eqnarray}
For $x = 2$, we obtain
\begin{eqnarray} f(2) = \frac{9}{2}, f'(2) = \frac{15}{4}.\end{eqnarray}

Now we proceed with the same function and use the operations defined by the differential algebra.
For $x$ in
\begin{eqnarray} f(x) = x^{2} + \frac{1}{x}, \end{eqnarray}
we substitute $x \rightarrow (x, 1) = (2, 1)$ and use our rules:
\begin{eqnarray}
f[(2,1)] &= &(2,1)^{2} + (2,1)^{-1}\\
         &= &(4,4) + \left(\frac{1}{2}, -\frac{1}{4}\right)\\
         &= &\left(\frac{9}{2}, \frac{15}{4}\right) = (f(2), f'(2)).
\end{eqnarray}
The computation of derivatives becomes an algebraic problem, no need for small numbers, it is exact.

\subsection{Differential algebra: higher orders and more variables}
\subsubsection{Higher orders}
The algorithm can be extended to higher order, let us say of order $N$:
\begin{enumerate}
\item[1.] the pair ($q_{0}, 1$) becomes a vector of order $N$,
$(q_{0}, 1) \rightarrow (q_{0}, 1, 0, 0, \ldots, 0)$,
$\delta = (0, 1, 0, 0, 0, \ldots)$;

\item[2.] $(q_{0}, q_{1}, q_{2}, \ldots, q_{N}) + (r_{0}, r_{1}, r_{2}, \ldots, r_{N})
= (s_{0}, s_{1}, s_{2}, \ldots, s_{N}),$ with $s_{i} = q_{i} + r_{i}$;

\item[3.] $c \cdot (q_{0}, q_{1}, q_{2}, \ldots, q_{N}) = (c\cdot q_{0},c\cdot q_{1},
c\cdot q_{2}, \ldots, c\cdot q_{N})$;

\item[4.] $(q_{0}, q_{1}, q_{2}, \ldots, q_{N}) \cdot (r_{0}, r_{1}, r_{2}, \ldots, r_{N}) = (s_{0}, s_{1}, s_{2}, \ldots, s_{N});$
with
\begin{eqnarray}s_{i} = \sum_{k=0}^{i} \frac{i!}{k! (i-k)!} q_{k} r_{i-k}.
\end{eqnarray}
\end{enumerate}

\subsubsection{Additional variables}
We can go to more dimensions, i.e.\ more variables, with rather small modifications.
If we had started with
\begin{eqnarray} x = (a,1,0,0,0,\ldots ), \end{eqnarray}
we would get
\begin{eqnarray}
f(x) = (f(a), f'(a), f''(a), f'''(a), \ldots, f^{(n)}(a)). \end{eqnarray}
This can be extended to more variables $x$,~$y$:
\begin{eqnarray}
x = (a,1,0,0,0,\ldots), \quad {\rm d}x = (0,1,0,0,0,\ldots ),
\end{eqnarray}
\begin{eqnarray}
y = (b,0,1,0,0,\ldots), \quad {\rm d}y = (0,0,1,0,0,\ldots )
\end{eqnarray}
and we get (with more complicated multiplication rules~\cite{bib:forest, bib:ef05})
\begin{eqnarray}
f((x+{\rm d}x), y+{\rm d}y)) = \left(f, \frac{\partial f}{\partial x},
\frac{\partial f}{\partial y}, \frac{\partial^{2} f}{\partial x^{2}},
\frac{\partial^{2} f}{\partial x \partial y},\ldots  \right)(x, y). \end{eqnarray}

\subsection{Examples}
To demonstrate the algorithm and how it can be used in practice, I show
a few examples.

\subsubsection{Simple example}
Below is a very simple program to compute $\sin(\pi/6)$.
In the first case (DATEST1) regular data types are used while in the second
program (DATEST2) the usual data types are replaced by the type of a
Taylor map (in the program called {\bf{my\_taylor}}).
~~\\
~~\\
~~\\
\begin{minipage}[b]{0.5\textwidth}
PROGRAM DATEST1\\
{{use my\_own\_da}}\\
{\bf{real(8)  x,z,dx}}\\
my\_order=3\\
dx=0.0        \\
x=3.141592653/6.0 + dx\\
call track(x, z)\\
call print(z, 6)\\
END PROGRAM DATEST1\\
~~\\
SUBROUTINE TRACK(a, b)\\
{{use my\_own\_da}}\\
{\bf{real(8)  a,b}}\\
b = sin(a)\\
END SUBROUTINE TRACK\\
\end{minipage}%
\begin{minipage}[b]{0.5\textwidth}
PROGRAM DATEST2   \\
{{use my\_own\_da}}\\
{\bf{type(my\_taylor) x,z,dx}}\\
my\_order=3\\
dx=1.0.mono.1~~~~!{\tiny{this is our (0,1)}}\\
x=3.141592653/6.0 + dx\\
call track(x, z)\\
call print(z, 6)\\
END PROGRAM DATEST2 \\
~~\\
SUBROUTINE TRACK(a, b)\\
{{use my\_own\_da}}\\
{\bf{type(my\_taylor) a,b}}\\
b = sin(a)\\
END SUBROUTINE TRACK\\
\end{minipage}
If we run the two programs, we obtain for the two outputs\\
~~~\\
~~~\\
\begin{minipage}[b]{0.5\textwidth}
    {{(0,0)  0.5000000000000E+00}}\\
    ~~~~\\
    ~~~~\\
    ~~~~\\
    ~~~~\\
    ~~~~\\
    ~~~~\\
    ~~~~\\
    ~~~~\\
    ~~~~\\
\end{minipage}%
\begin{minipage}[b]{0.5\textwidth}
    {{(0,0)  0.5000000000000E+00}}\\
    (1,0)  0.8660254037844E+00\\
    (0,1)  0.0000000000000E+00\\
    (2,0) -0.2500000000000E+00\\
    (0,2)  0.0000000000000E+00\\
    (1,1)  0.0000000000000E+00\\
    (3,0) -0.1443375672974E+00\\
    (0,3)  0.0000000000000E+00\\
    (2,1)  0.0000000000000E+00\\
    (1,2)  0.0000000000000E+00\\
\end{minipage}
We have $\sin({\pi}/{6}) = 0.5$ all right, but what is the rest?
Let us look at
\begin{eqnarray}
\sin\left(\frac{\pi}{6} + \Delta x\right) = \sin\left(\frac{\pi}{6}\right)
+ \cos\left(\frac{\pi}{6}\right) \Delta x^{1} - \frac{1}{2}\sin\left(\frac{\pi}{6}\right)
\Delta x^{2} - \frac{1}{6}\cos\left(\frac{\pi}{6}\right) \Delta x^{3},
\end{eqnarray}
i.e.\ we have found the coefficients of the $sine$ evaluated at $\sin({\pi}/{6})$.
\begin{itemize}
\item[i)] We have used a simple algorithm here ($\sin$) but it can be {{anything}} and very complex.
\item[ii)] We can compute non-linear maps as a Taylor expansion of {{anything}} the program computes.
\item[iii)] Simply by:
\begin{itemize}
\item[a)] replacing regular (e.g. real) types by TPSA types (my\_taylor), i.e.\ variables $x, p$ are automatically replaced by $(x,1,0,\ldots )$ and $(p,0,1,0,\ldots )$ etc.;
\item[b)] operators and functions ($+, -, *, =, \ldots,  \exp, \sin, \ldots$)
are automatically overloaded, i.e.\ behave according to the new type.
\end{itemize}
\end{itemize}

\subsubsection{A FODO lattice}
Another example.
\begin{itemize}
\item[$\bullet$] Track through a simple FODO lattice described as:
\item[] QF - DRIFT - QD
\item[] Integrate 100 steps in the quadrupoles.
\item[] Now we use {{three}} variables:
\item[] x, p, $\Delta p = (z(1), z(2), z(3))$
(third variable needed to compute the chromaticity).
\end{itemize}
~~~\\
~~~\\
~~~\\
\begin{minipage}[b]{0.5\textwidth}
PROGRAM FODO1\\
{{use my\_own\_da}}\\
{{use my\_analysis}}\\
type(my\_taylor) z(3)\\
{{type(normalform) NORMAL}}\\
{{type(my\_map) M,id}}\\
~~\\
real(dp) L,DL,k1,k3,fix(3) \\
integer i,nstep\\
~~\\
~~\\
my\_order=4~~! maximum order 4 \\
fix=0.0  ! fixed point\\
id=1\\
z=fix+id\\
~~\\
LC=62.5~~~~~! half cell length\\
L=3.0~~~~~! quadrupole length\\
nstep=100\\
DL=L/nstep\\
k1=0.003~~~~~! quadrupole strength\\
{{~~~~~~~}}\\
\end{minipage}%
\begin{minipage}[b]{0.5\textwidth}
do i=1,nstep~~~~~! track through quadrupole\\
z(1)=z(1)+DL/2*z(2)\\
z(2)=z(2)-k1*DL*z(1)/(1 + z(3))\\
z(1)=z(1)+DL/2*z(2)\\
enddo\\
z(1)=z(1)+LC*z(2)~~~~~! drift of half cell length\\
\\
do i=1,nstep~~~~~! track through quadrupole\\
z(1)=z(1)+DL/2*z(2)\\
z(2)=z(2)-k1*DL*z(1)/(1 + z(3))\\
z(1)=z(1)+DL/2*z(2)\\
enddo\\
{{~~~~~~~~~~~~~~~~~~~~~~~~~~~~~~}}\\
z(1)=z(1)+LC*z(2)~~~~~! drift of half cell length\\
\\
{{call print(z(1),6)}}\\
{{call print(z(2),6)}}\\
{{M=z}}  \\
{{NORMAL=M}}\\
write(6,*) {{normal\%tune, normal\%dtune\_da}}\\
end program fodo1\\
\end{minipage}
~~~\\
\clearpage
We get the result of the program:\\
~~~\\
~~~\\
\begin{minipage}[b]{0.5\textwidth}
   (0,0,0) {{0.9442511679729E-01}}\\
   (0,0,1) {{-0.9729519276183E-01}}\\
~~~~~~~~~~~~~~~~~~~~~~~~~~~~~~~\\
   (1,0,0) {{0.6972061935061E-01}}\\
   (0,1,0) {{0.1677727932585E+03}}\\
   (1,0,1) 0.1266775134236E+01\\
   (0,1,1)-0.3643444875882E+02\\
   (1,0,2)-0.1603248617779E+01\\
   (0,1,2) 0.3609522079691E+02\\
   (1,0,3) 0.1939697138318E+01\\
   (0,1,3)-0.3575511053483E+02\\
~~~~~~~~~~~~~~~~~~~~~~~~~~~~~~~\\
   (1,0,0){{-0.5300319873866E-02}}\\
   (0,1,0) {{0.1588490329398E+01}}\\
   (1,0,1) 0.1060055415702E-01\\
   (0,1,1)-0.5832024543075E+00\\
   (1,0,2)-0.1590066005419E-01\\
   (0,1,2) 0.5779004431627E+00\\
   (1,0,3) 0.2120059477024E-01\\
   (0,1,3)-0.5725843143370E+00\\
\end{minipage}%
\begin{minipage}[b]{0.5\textwidth}
~~~\\
With only linear elements in the Taylor expansion,
the result for the matrix per cell is
~~~\\
\begin{eqnarray}
\Delta x_{f} = {{0.06972}} \Delta x_{i} + {{167.77}} \Delta p_{i},
\end{eqnarray}
\begin{eqnarray}
\Delta p_{f} = {{-0.00530}} \Delta x_{i} + {{1.5885}} \Delta p_{i}.
\end{eqnarray}
~~~\\
The outputs from the normal-form analysis are (per cell) \\
Tune = {{0.094425}}, \\
Chromaticity =  {{--0.097295}.} \\
~~~\\
\end{minipage}
~~\\
\clearpage
We add now to this example a thin octupole:\\
~~\\
~~\\
~~\\
\begin{minipage}[b]{0.5\textwidth}
PROGRAM FODO3\\
{{use my\_own\_da}}\\
{{use my\_analysis}}\\
type(my\_taylor) z(3)\\
{{type(normalform) NORMAL}}\\
{{type(my\_map) M,id}}\\
~~\\
real(dp) L,DL,k1,k3,fix(3) \\
integer i,nstep\\
~~\\
~~\\
my\_order=4~~! maximum order 4 \\
fix=0.0  ! fixed point\\
id=1\\
z=fix+id\\
~~\\
LC=62.5~~~~~! half cell length\\
L=3.0~~~~~! quadrupole length\\
nstep=100\\
DL=L/nstep\\
k1=0.003~~~~~! quadrupole strength\\
{{k3=0.01}}~~~~~! octupole strength\\
{{~~~~~~~}}\\
\end{minipage}%
\begin{minipage}[b]{0.5\textwidth}
do i=1,nstep~~~~~! track through quadrupole\\
z(1)=z(1)+DL/2*z(2)\\
z(2)=z(2)-k1*DL*z(1)/(1 + z(3))\\
z(1)=z(1)+DL/2*z(2)\\
enddo\\
{{z(2)=z(2)-k3*z(1)**3/(1 + z(3))}}~~! octupole kick\\
z(1)=z(1)+LC*z(2)~~~~~! drift of half cell length\\
\\
do i=1,nstep~~~~~! track through quadrupole\\
z(1)=z(1)+DL/2*z(2)\\
z(2)=z(2)-k1*DL*z(1)/(1 + z(3))\\
z(1)=z(1)+DL/2*z(2)\\
enddo\\
{{~~~~~~~~~~~~~~~~~~~~~~~~~~~~~~}}\\
z(1)=z(1)+LC*z(2)~~~~~! drift of half cell length\\
\\
{{call print(z(1),6)}}\\
{{call print(z(2),6)}}\\
{{M=z}}  \\
{{NORMAL=M}}\\
write(6,*) {{normal\%tune, normal\%dtune\_da}}\\
end program fodo3\\
\end{minipage}
~~~\\
\clearpage
The result of this program is now\\

\begin{minipage}[b]{0.5\textwidth}
~~~~~~~~~~~~~~~~~~~~~~~~~~~~~~~\\
~~~~~~~~~~~~~~~~~~~~~~~~~~~~~~~\\
   (0,0,0) {{0.9442511679729E-01}}\\
   (0,0,1) {{-0.9729519276183E-01}}\\
~~~~~~~~~~~~~~~~~~~~~~~~~~~~~~~\\
   (2,0,0) {{0.5374370086899E+02}}\\
   (0,2,0) {{0.5374370086899E+02}}\\
   (0,0,2) 0.1018391758451E+00\\
   (2,0,1) 0.2035776281196E+02\\
       ..........              \\
~~~~~~~~~~~~~~~~~~~~~~~~~~~~~~~\\
   (1,0,0) {{0.6972061935061E-01}}\\
   (0,1,0) {{0.1677727932585E+03}}\\
   (1,0,1) 0.1266775134236E+01\\
   (0,1,1)-0.3643444875882E+02\\
   (3,0,0)-0.1586519461687E+01\\
   (2,1,0)-0.1440953324752E+02\\
   (1,2,0)-0.4362477179879E+02\\

       ..........              \\
   (1,0,0){{-0.5300319873866E-02}}\\
   (0,1,0) {{0.1588490329398E+01}}\\
   (1,0,1) 0.1060055415702E-01\\
   (0,1,1)-0.5832024543075E+00\\
   (3,0,0)-0.1519218878892E-01\\
\end{minipage}%
\begin{minipage}[b]{0.5\textwidth}
\small\bf
~~~\\
Now there are non-linear elements in the Taylor expansion;
the outputs from the normal-form analysis are (per cell) \\
Tune = {{0.094425}}, \\
Chromaticity =  {{$-$0.097295}}. \\
The detuning with amplitude is {{53.74}}.
This is the result we had for the analytical calculation for an octupole
in the previous section.
\end{minipage}

\subsubsection{Analysis using the normal form}
Remember the normal-form transformation
\begin{eqnarray}
{\cal{A}}{\cal{M}}{\cal{A}}^{-1} = {\cal{R}}.
\end{eqnarray}

The type {{normalform}} in the demonstration package also contains the maps ${\cal{A}}$ and ${\cal{R}}$:
\begin{center}
{{j2=(x**2+p**2)*NORMAL\%A**(-1)}}
\end{center}
Remember that x**2+p**2 is the tilted ellipse and
${\cal{A}}^{-1}$ the transformation to the normal form,
containing the optical parameters $\alpha$, $\beta$, $\gamma$.

We can obtain the optical functions because:
\begin{itemize}
\item[i)] $\beta$ is the coefficient of p**2 of invariant j2;
\item[ii)] $\alpha$ is the coefficient of x*p of invariant j2;
\item[iii)] $\gamma$ is the coefficient of x**2 of invariant j2.
\end{itemize}
To obtain the linear optical parameters in our code, use something like
\begin{center}
{{$\beta$ =  j2.sub.beta}},\\
{{$\alpha$ =  0.5*j2.sub.twoalpha}},\\
{{$\gamma$ =  j2.sub.gamma}};
\end{center}
we obtain (here at the end of the cell) $\beta = 300.080714$,
$\alpha = -1.358246$, $\gamma = ~9.480224$E-3.

\section{Summary and epilogue}
In this article
I have tried to show the need for a new paradigm and a new framework for
the study and analysis of non-linear beam dynamics is accelerators.
Emphasis was put on the use of new numerical and computational techniques
which allow better (and simpler) treatment.
Putting aside pedagogical issues, the dreadful use of approximations
and the limitations in the standard treatments can be avoided
in the era of powerful computers and computing techniques.
For example, the evaluation of resonances can be easily carried beyond
leading orders and brings us well ahead of standard models.
Therefore, I advocate the use of contemporary tools
which unify linear, non-linear, coupled, decoupled and symplectic
treatments. They can be treated and handled at the same time and the
theoretical framework is simplified in the humble opinion of the author.
The ideas presented in this article are based on the work of numerous
people and at this point I hope that the next generation of accelerator
physicists will continue (or start) to follow in this direction.
\section*{Acknowledgements}
I wish to thank my colleagues for their interest and many discussions
on this topic.
I am glad to thank the members of the SSC Central Design Group
whose contributions to the topics presented here have inspired me to
enter this subject.
I am particularly grateful to Prof.~Alex~Dragt who was the first to
apply the technique of Lie transformation to non-linear problems in
accelerator physics.
For their contributions and many discussions I appreciate the
collaboration with Dr~Dobrin~Kaltchev and Dr~Etienne~Forest,
the latter for the TPSA demonstration program used during the
preparation of this article.

\end{document}